\documentclass[iop,apj]{emulateapj}
\usepackage{times}
\usepackage{epsfig}
\usepackage{amsmath, amsthm, amssymb}
\usepackage{bm}
\usepackage[plainpages=false, colorlinks=true, anchorcolor=blue, linkcolor=blue, citecolor=blue, bookmarks=false]{hyperref}
\usepackage{color}
\usepackage{microtype}

\newcommand{\beq}{\begin{equation}}
\newcommand{\eeq}{\end{equation}}
\newcommand{\bea}{\begin{eqnarray}}
\newcommand{\eea}{\end{eqnarray}}

\newcommand{\bB}{\mathbf{B}}
\newcommand{\bn}{\mathbf{n}}
\newcommand{\bu}{\mathbf{u}}
\newcommand{\bE}{\mathbf{E}}
\newcommand{\bx}{\mathbf{x}}
\newcommand{\btimes}{\pmb{\times}}

\slugcomment{Submitted to the Astrophysical Journal}

\begin{document}

\title{The Biermann Catastrophe in Numerical MHD}
\author{Carlo Graziani, Petros Tzeferacos, Dongwook Lee, Donald Q. Lamb, Klaus Weide,
Milad Fatenejad, \& Joshua Miller}

\affil{Flash Center for Computational Science, Department of Astronomy
\& Astrophysics, University of Chicago, Chicago, IL, 60637}

\email{carlo@oddjob.uchicago.edu}

\shorttitle{THE BIERMANN CATASTROPHE}
\shortauthors{GRAZIANI ET AL.}

\begin{abstract}

The Biermann Battery effect is frequently invoked in 
cosmic magnetogenesis and studied in High-Energy Density laboratory physics 
experiments.  Generation of magnetic fields by the Biermann effect due to 
mis-aligned density and temperature gradients in smooth flow \textit{behind} 
shocks is well known.  We show that a Biermann-effect magnetic field is also generated 
\textit{within} shocks.
Direct implementation of the Biermann effect in MHD 
codes does not capture this physical process, and worse, produces unphysical 
magnetic fields at shocks whose value does not converge with resolution.  We 
show that this convergence breakdown is due to naive discretization, which fails
to account for the fact that discretized irrotational
vector fields have spurious solenoidal components that grow without bound near a discontinuity. We 
show that careful consideration of the kinetics of ion viscous 
shocks leads to a formulation of the Biermann effect that gives rise 
to a convergent algorithm.  We note two novel physical effects:
a \textit{resistive magnetic 
precursor} in which Biermann-generated field in the 
shock ``leaks'' resistively upstream; and a \textit{thermal 
magnetic precursor}, in which field is generated by the Biermann 
effect ahead of the shock front due to gradients created by the shock's 
electron thermal conduction precursor.  Both effects appear to be 
potentially observable in experiments at laser facilities.  We re-examine published
studies of magnetogenesis in galaxy cluster formation, 
and conclude that the simulations in question
had inadequate resolution to reliably estimate the field generation rate.  
Corrected estimates suggest primordial field values in the range 
$B\sim 10^{-22}$G --- $10^{-19}$G by $z=3$.

\keywords{magnetohydrodynamics --- plasmas --- magnetic fields}

\end{abstract}

\section{Introduction}

Dynamo theories of proto- and extra-galactic primordial magnetic fields, 
which endeavor to explain how those fields achieved their current strength 
and structure, work by amplifying small initial seed fields by means of 
turbulent plasma motions \citep{Kronberg_1994,Kulsrud_Zweibel_2008}.  
However, the induction equation of resistive magnetohydrodynamics (MHD),
\beq
\frac{\partial\bB}{\partial t} = 
\nabla\btimes
\left\{
\bu\btimes\bB
-\frac{\eta c^2}{4\pi}\nabla\btimes\bB
\right\},
\label{eq:Induction_Equation}
\eeq
always admits the solution $\bB(\bx,t)=0$. This simple 
observation poses a problem for the generation of the required seed fields, 
as they cannot be created in ideal MHD starting from a field-free state.

There have been several proposals for generating the required seed fields from
mechanisms such as primordial phase transitions, or from processes occurring during
inflation \citep[see reviews in][]{Widrow2002,Widrow2012}.
The Biermann battery effect \citep{biermann1950} provides another popular 
resolution of this problem \citep {Kulsrud_1997,Widrow2002,Kulsrud_Zweibel_2008}. The 
effect, which arises in consequence of the large difference in the electron 
and ion mass, is attributable to small-scale charge separation in the 
plasma.  Pressure forces produce much larger accelerations of electrons than 
of ions, and the relative acceleration of the two components results in 
charge separation that must be balanced by an electric field 
\beq
\mathbf{E}_B\equiv -(en_e)^{-1}\nabla P_e,
\label{eq:Biermann_E}
\eeq
where $n_e$ and $P_e$ are the electron
number density and pressure, respectively.  Since this field is not, in 
general, irrotational, it can act as a source of magnetic field in the 
induction equation, 
\beq
\frac{\partial\bB}{\partial t} = 
\nabla\btimes
\left\{
\bu\btimes\bB
-\frac{\eta c^2}{4\pi}\nabla\btimes\bB
+\frac{c}{en_e}\nabla P_e
\right\},
\label{eq:Induction_Equation_Biermann}
\eeq
generating a non-zero $\bB$ from an initially unmagnetized state.

The Biermann battery effect has been successfully invoked in numerical 
simulations exploring the generation of seed fields in cosmological 
ionization fronts \citep{subramanian1994,Gnedin_2000}, protogalaxies \citep 
{Davies_Widrow_2000}, and Pop-III star formation \citep{xu_2008}. Magnetic 
field generation by the Biermann mechanism is also of significant interest 
in direct-drive and indirect-drive inertial confinement fusion, where 
strong gradients behind the converging shock can lead to dynamically important field 
strengths \citep{srinivasan2013}, and more generally in the field of 
High-Energy Density Physics, where the effects of field generation \citep{gregori2012} and 
amplification \citep{meinecke2014} can be examined in a laboratory setting at laser 
facilities, in experiments where large gradients are produced in strong 
plasma shocks \citep{fryxell2010,tzeferacos2012,tzeferacos2014}.

While the generation of magnetic fields by the Biermann effect as a result 
of strong mis-aligned density and temperature gradients in the smooth flow 
{\em behind} shocks is well known, we show that there exists a previously 
unrecognized Biermann effect due to the electron-ion charge separation that
occurs {\em within} ion viscous shocks. This
effect is a consequence of the kinetic theory of shock structure in plasmas, as
we show below.

A straightforward implementation of the Biermann effect in finite-volume 
Eulerian, purely Langrangian, and ALE codes, whether as a split source term 
or as a flux term, does not capture this physical process, and worse, leads 
to non-convergent results \citep{Fatenejad2013}.  In symmetric situations 
such as planar or spherical shocks, where no field should arise, such 
implementations produce anomalous field generation that grows without bound 
with resolution \citep{Fatenejad2013}. This behavior is observed across a 
range of different MHD codes \citep[see the discussion of codes in][] 
{Fatenejad2013}.  This is the Biermann catastrophe of numerical MHD. We show 
that this failure is intimately related to the failure of such codes to 
correctly model the structure of the plasma shock.

In a gasdynamic/MHD formulation, where the shocks are modeled as zero-width discontinuities
of the flow, the trouble arises from the behavior of the Biermann flux, which is to say 
from the electric field, Eq.~(\ref {eq:Biermann_E}), in the vicinity of a
shock. The gradient $\nabla P_e$, which analytically speaking 
acquires a Dirac $\delta$ component at the shock, is ascribed a numerical 
magnitude that grows without bound at the shock with increasing resolution. 
It is this divergence that is connected with failure of MHD codes 
to correctly predict shock-driven magnetic field generation in supposedly 
simple test cases, where the correct value of the generated field is zero.
The investigation of this failure is one of the central concerns 
of this article. 

That this failure was not previously recognized is a consequence of the 
history of the numerical study of the Biermann Battery effect as a 
source of cosmic seed fields, which has a curious feature with respect 
to shocks. On the one hand, the importance of shocks to lifting barotropic 
constraints, thus making available the kind of non-aligned gradients 
required to drive the Biermann effect, has been widely recognized
\citep{Kulsrud_1997,Davies_Widrow_2000}.  On the other hand, the direct 
effect of shocks -- as opposed to that of their trailing downstream 
gradients -- on this magnetogenesis has not really been carefully examined.  
This is probably because the computational strategies that have been adopted 
both circumvent difficulties at shocks and direct attention away from the 
magnetizing properties of shocks.  For example, \citet{Kulsrud_1997} and 
\citet{Gnedin_2000} perform essentially hydrodynamic simulations, 
in which magnetic fields evolved by the induction equation have no dynamical role. 
\citet{Davies_Widrow_2000} forgo the induction equation altogether, in favor
of the equation of inviscous evolution of vorticity $\bm{\omega}$
\beq
\frac{\partial\bm{\omega}}{\partial t}=
\nabla\btimes
\left\{
\bu\btimes\bm{\omega}
+\frac{1}{\rho}\nabla P
\right\},
\label{eq:vorticity}
\eeq
which, by comparison with Eq.~(\ref{eq:Induction_Equation_Biermann}), implies
that inviscous, non-resistive plasmas satisfy the equation
\beq
\frac{\partial\bB}{\partial t}=
\frac{cM_i}{e(1+\bar{Z})}\frac{\partial\bm{\omega}}{\partial t}.
\label{eq:Mag_vs_Vorticity}
\eeq
\citep{Kulsrud_Zweibel_2008}.  On its face, this relation would appear to suggest that in an initially
unmagnetized and irrotational plasma, the magnetic field is simply proportional
to the vorticity, $\bB=\frac{cM_i}{e(1+\bar{Z})}\bm{\omega}$, where $M_i$ is the
ion mass and $\bar{Z}$ is the average ionization fraction.

One difficulty with these approaches is that they entirely 
neglect to treat the field generation \textit{within the shock itself}, as we noted 
above.  There are indeed large, non-aligned
gradients at the simulated shocks, but these gradients are unphysical side-effects
of the numerical strategies used to integrate the hydrodynamic equations, and consequently are 
resolution-dependent.  It is therefore hopeless to expect convergence with resolution of the resulting magnetic 
fields.

Furthermore, any magnetization generated by the shock itself imprints itself 
on the magnetic field structure as the shock moves through, leaving behind a 
substantial residue that is superposed on the smooth-flow Biermann-generated 
field.  It is essential to come to a correct understanding of the behavior 
of the Biermann term at shocks, to have any confidence that results arrived 
at in this manner bear any resemblance to reality.  In particular, the 
presumption that shock jump condition on vorticity should bear a relation to 
the jump condition on magnetization that preserves the proportionality 
between the two has been hypothesized but never demonstrated \citep 
{Kulsrud_1997,Davies_Widrow_2000}, and seems in fact farfetched: it would 
imply that magnetization \textit{is} vorticity, that is, that the 
coincidence expressed by Eq.~(\ref{eq:Mag_vs_Vorticity}) in fact reflects a 
deep identity, and that magnetic degrees of freedom of ideal MHD are somehow 
already contained in unmagnetized Eulerian hydrodynamics, encoded in 
derivatives of the velocity field.  Such a claim is difficult to accept, and to 
even attempt to support it would require an analysis of the modification 
brought by the Biermann Battery term to the Rankine-Hugoniot jump condition 
on $\bB$, an analysis that we furnish for the first time in this paper.  We 
will return to a discussion of vorticity in \S\ref{sec:discussion}.

There have also been full MHD simulations of magnetogenesis through the
Biermann effect in galaxy clusters\citep{xu_2008}, which were performed using the ENZO code
\citep{xu2010,xu2011,Bryan_2014}. No convergence study of Biermann-generated
magnetic fields was reported for these simulations,
possibly because of the paucity of non-trivial analytical verification tests
against which the code's implementation of the Biermann effect could be tested.

It is clearly long past time that the behavior of the Biermann effect at an 
MHD shock be fully analyzed and characterized, so as to furnish a 
mathematical-physics target at which algorithmic implementations can shoot. 
In order to do so, it is necessary to draw connections between the 
well-understood kinetic theory of planar plasma shocks \citep
{Shafranov_Plasma_Shock_1957,jaffrin1964,zeldovich_raizer} and 
that of spatially-inhomogeneous MHD shocks, connections which 
have not so far been made or exploited in the literature.

In particular, it is essential to address the following questions:
 
\begin{enumerate} 

\item How does the kinetic theory of the structure of plasma shocks inform our 
understanding of the Biermann effect in shocks, and how can it guide us to 
an accurate and convergent treatment of the effect in MHD codes?

\item Do we have any right to expect convergence with resolution from an MHD 
simulation with a source term as ostensibly misbehaved at shocks as that of 
Eq.~(\ref {eq:Biermann_E})?  In other words, is the Biermann source term
mathematically well-defined near a discontinuity of a weak solution of the MHD
equations, and, if so, how does it affect the Rankine-Hugoniot jump condition
on $\bB$?

\item Assuming the Biermann source term, Eq.~(\ref{eq:Biermann_E}) can be 
interpreted in a sensible manner in the vicinity of an inviscid shock, are 
its predictions consistent with the predictions of kinetic theory near a 
shock? Or does the Biermann term in MHD need to be flux-limited, in the 
style of thermal and radiation source terms whose misbehavior must be 
limited on kinetic theory grounds when gradients grow too large?


\end{enumerate}

We address question (1) in \S\ref{sec:plasma_shocks}, abstracting the
essential ingredients of plasma shock theory required to construct a
valid MHD model of the Biermann effect due to shocks; whereas we address the
remaining two questions in \S\ref{sec:biermann_at_shocks}, where we 
demonstrate that (2) in fact the Biermann source  of Eq.~(\ref 
{eq:Biermann_E}) is mathematically consistent and well-behaved near 
weak-solution discontinuities, and (3) the Biermann source term in fact 
yields the correct EMF across the shock, matching the prediction of plasma 
shock theory \citep{zeldovich_raizer,amendt2009,jaffrin1964}.

We clarify the origin of the Biermann catastrophe as a numerical effect 
attributable to the difficulty of discretizing the source term of Eq.~(\ref
{eq:Biermann_E}) in the vicinity of a shock.  We show that the numerical 
anomaly can be eliminated by leveraging the continuity of the electron 
temperature $T_e$ across shocks -- a benefit of the kinetic-theory 
connection.  Reformulation of the Biermann source term in terms of $T_e$ 
allows the singularity to be isolated, and the flux of magnetic field due to 
the Biermann effect to be rewritten in a manifestly finite
form suitable for translation into a convergent numerical algorithm.   

The Biermann effect is due to electron-ion charge separation, and is 
sensitive to departure from thermal equilibrium between electrons and ions. 
Such a departure is precisely what occurs at shocks, so that a correct 
treatment of the effect at shocks necessarily requires that the 
disequilibrium be modeled. For this, a 2-temperature plasma model is 
mandatory.  An interesting consequence of this observation is the fact that 
the Biermann effect is enhanced not only at shocks, but also at contact 
discontinuities, at ionization fronts, and at material species boundaries, 
because the electron partial pressure is discontinuous at such surfaces, 
even though the total pressure is continuous there.  This enhancement cannot 
be computed in an equilibrium treatment of the plasma.

An additional point of this paper is to point out two novel and interesting 
effects associated with the Biermann effect in the neighborhood of a shock. 
In \S\ref{subsec:resistive_precursor}, we show that a \textit{resistive 
magnetic precursor} is generated in resistive MHD, wherein magnetic field 
generated by the Biermann effect in the shock ``leaks'' resistively from the 
shock into a region of the upstream fluid whose physical extent is 
proportional to the resistivity. And in \S\ref
{subsec:thermal_magnetic_precursor}, we show that a \textit{thermal magnetic 
precursor} is generated ahead of the shock through the Biermann effect by 
plasma motions generated by the shock's electron thermal conduction 
precursor.  Both effects are potentially observable in laboratory conditions 
at high-intensity laser facilities such as Vulcan, Omega, and NIF.  We 
discuss the observability of these effects at laser facilities.  We show 
that appropriately-designed experiments at such facilities could currently 
observe the resistive magnetic precursor, providing a clean experimental 
validation test of the Biermann effect in plasma shocks.  We also argue that 
the smaller thermal magnetic precursor might become observable in future 
experiments.

\section{Review of Kinetic Theory of Plasma Shocks}\label{sec:plasma_shocks}

We begin by reviewing some essential results from the kinetic theory of shocks
in plasmas.  The basic theory of the fluid structure of planar shocks in plasmas
was set out in \citet{Shafranov_Plasma_Shock_1957}, while the electromagnetic
structure of such shocks was discussed in \citet{jaffrin1964}, and more recently
in \citet{amendt2009}.  An extremely lucid presentation of these results may
be found in Chapter VII of \citet{zeldovich_raizer}.

There are three essential ingredients to be imported from the kinetic theory 
of plasma shocks in order to fashion a working MHD model of the Biermann 
effect: the loss of thermal equilibrium between electrons and ions at 
shocks; the adiabatic behavior of electrons, up to electron heat conduction; 
and the charge-separation-induced electric field across the shock front.  We
now review these in order, and conclude by presenting the MHD model that forms
the basis for the rest of this paper.

\subsection{Ion-Electron Disequilibrium}

As discussed on p.36 of \citet{drake2006}, a strong shock disturbs the 
thermal equilibrium between electrons and ions in a plasma.  That 
equilibrium is maintained by electron-ion collisions, and operates over 
timescales $\tau_{ei}$ that are long compared to the shock-crossing time of 
a parcel of fluid entering the ion viscous shock in consequence of the large 
ratio $m_i/m_e$  \citep{spitzer1962}.  As a result, it is essential to 
describe the fluid in terms of an additional degree of freedom -- the 
electron temperature, $T_e$ -- with respect to the usual equilibrium MHD 
model.

Fortunately, it is unnecessary to model the fluid using 
new inertial degrees of freedom to describe the electron fluid.  A single 
inertial component for the fluid as a whole gives an adequate description of the fluid 
structure near the shock, and a completely satisfactory description of the 
fluid in smooth flow regions, where electron-ion collisions restore local 
thermal equilibrium \citep{drake2006, zeldovich_raizer}.  This makes it 
easier to adapt existing MHD codes to treat the Biermann effect correctly, 
since it is much easier to add a scalar degree of freedom for $T_e$ than it 
would be to deal with two velocity fields, one each for the electron fluid 
and for the ion fluid.

\subsection{Nearly Adiabatic Electrons}\label{subsec:thermal_structure}

The large ion-to-electron mass ratio, which we recall from the discussion 
preceding Eq.~(\ref{eq:Biermann_E}) is responsible for the charge separation 
that produces the Biermann effect, has the further consequence that electron 
mobility is much higher than ion mobility, and that thermal conductivity due 
to electron-electron collisions dominates the heat transport in the fluid. 
At the same time, the forces between electrons and ions as the fluid crosses 
the ion viscous shock front are effectively dissipation-free, as they are 
simply electrostatic fields generated by charge separation, and collisional 
dissipation processes are too slow to act during the shock-crossing timescale.

These observations lead to the conclusion that while the ions undergoing 
shock compression experience the usual thermodynamically-irreversible, 
entropy-generating process associated with shocks, the electrons are 
compressed \textit{adiabatically} by the electrostatic forces exerted upon 
them by the ions, and consequently do not suffer entropy increments due to 
shock compression.  The electron entropy would be a passively-advected 
scalar, then, except for the dissipative effect of electron thermal 
conduction, and for the slow (relative to timescales relevant to the shock) 
effect of electron-ion collisional equilibration.  Electron thermal 
conduction effectively rules out any sudden change in $T_e$ at the shock, 
since such a change would produce an enormous restoring heat flux to heal 
the discontinuity.

The fluid structure that follows from these considerations is of a sudden 
discontinuous compression of the ions at the shock, accompanied by a smooth 
increase in $T_e$, which is continuous throughout the shock (in the Eulerian 
limit where the shock width tends to zero, $T_e$ acquires a discontinuous 
derivative in the direction of the shock normal).  The electron temperature 
exhibits a \textit{thermal precursor} that leads the shock.  The size 
$\lambda_T$ of the precursor region may be calculated by balancing advection 
against heat diffusion in the frame of the shock, and is about 
\begin{equation} 
\lambda_T\approx \frac{\kappa_e}{\rho c_{v,e} D}, 
\label{eq:Lambda_T} 
\end{equation} 
where $\kappa_e$ is the electron thermal conductivity, $c_{v,e}$ is the 
electron specific heat at constant volume per unit mass, and $D$ is the 
shock speed \citep{Shafranov_Plasma_Shock_1957,zeldovich_raizer}.  The effect
of this precursor region on accretion shocks in galaxy clusters has been recently
studied in \citet{Smith2013}.

\subsection{Electric Structure}\label{subsec:Electric}

The sudden change in the motion of ions entering the shock sheath, together 
with the more highly mobile motion of the electrons in the vicinity of the 
shock, results in a zone of charge separation-driven electric field in the 
normal direction to the shock \citep{jaffrin1964, amendt2009, 
zeldovich_raizer}.  By solving the Navier-Stokes equation across the shock,
(treating the results as means of kinetic-theory distributions, since a fluid
picture is clearly invalid inside the viscous shock sheath), \citet{jaffrin1964}
calculated the full electric structure across the shock.  We do not require
the full structure here, since in the end we wish to adopt an Eulerian picture
of the shock, in which the width of the shock is zero.  For our purposes, the
EMF across the shock is all that is required.  It is
shown in \citet{amendt2009} that the EMF $\cal{E}$ is given by
\bea
{\cal E}&=&k_BT_e\ln\left(\frac{n_{e,d}}{n_{e,u}}\right)\nonumber\\
&=&k_BT_e\ln\left(\frac{\rho_d}{\rho_u}\right),
\label{eq:EMF}
\eea
where the subscripts ``$u$'' and ``$d$'' denote ``upstream'' and ``downstream'',
respectively, and where we've assumed complete ionization to obtain the
second line.

This electric field is due to charge separation, and thus has a common 
ancestry with the electric field $\bE_B$ in Eq.~(\ref{eq:Biermann_E}). 
However, it is proper to the shock, rather than to the smooth-flow portion 
of the fluid.  It is essential, for the purpose of physical consistency, to 
demonstrate that an MHD model implementing the Biermann effect should 
reproduce the shock-crossing EMF given in Eq.~(\ref{eq:EMF}).  That this is
possible is demonstrated in the next section.

\subsection{The MHD Model}

We wrap up this section by exhibiting the Biermann-effect-laden MHD model
implied by these kinetic-theory considerations.  The dynamical system describing
the model is given in conservation form as follows:
\begin{eqnarray}
&&\frac{\partial\rho}{\partial t}+\nabla\cdot\left(\rho\bu\right)=0\label{eq:Masscons}\\
&&\frac{\partial\rho\bu}{\partial t}
+\nabla\cdot\left[\rho\bu\bu-\frac{1}{4\pi}\bB\bB
+\mathbf{1}\left(P+\frac{\bB^2}{8\pi}\right)\right]=0\label{eq:Momcons}\\
&&\frac{\partial}{\partial t}\left[\rho\left(\frac{1}{2}\bu^2+\epsilon_T\right)+\frac{\bB^2}{8\pi}\right]
+\nabla\cdot\bigg[
\rho\bu\left(\frac{1}{2}\bu^2+\epsilon_T+\frac{P}{\rho}\right)+\nonumber\\
&&\hspace{0.2cm}\frac{c}{4\pi}\left(-(\bu/c)\btimes\bB+
\frac{c\eta}{4\pi}\nabla\btimes\bB
+\bE_B\right)\btimes\bB\bigg]=0\label{eq:Econs}\\
&&\frac{\partial\rho s_e}{\partial t}
+\nabla\cdot\left(\rho\bu s_e\right)=
-{T_e}^{-1}\nabla\cdot\left(-\kappa_e\nabla T_e\right)\nonumber\\
&&\hspace{3.0cm}+\frac{\rho c_{v,e}}{T_e\tau_{ei}}(T_i-T_e)\label{eq:Secons}\\
&&\frac{\partial\bB}{\partial t}
+\nabla\cdot\bigg[
\bu\bB-\bB\bu -\frac{c^2\eta}{4\pi}\nabla\bB
\bigg]
+c\nabla\btimes\bE_B=0.\label{eq:Induction}
\end{eqnarray}
In these 
equations, $P=P_e+P_i$ is the total material pressure, $\epsilon_T$ is
the total specific thermal energy per unit mass, and $s_e$ is the specific electron
entropy per unit mass.

Eqs.~(\ref{eq:Masscons}-\ref{eq:Momcons}) are the usual MHD equations of 
mass and momentum conservation.  The energy conservation equation, Eq.~(\ref 
{eq:Econs}) includes a term to account for resistive dissipation, and 
another to account for the Biermann effect.  

Eq.~(\ref{eq:Secons}) expresses the conservation of electron entropy $s_e$ 
(up to heat conduction and electron-ion heat exchange), which, as we pointed 
out above, is required by the same approximation $m_e/m_i\rightarrow 0$ that 
gives rise to the Biermann effect in the first place.  This equation is 
introduced to describe the additional degree of freedom required by the 
2-temperature plasma treatment.  

Finally, in the induction equation, Eq.~(\ref {eq:Induction}), the Biermann 
effect is expressed here as a curl, rather than in the conservation form -- 
a divergence of a flux tensor -- that it will assume later. 

These dynamical equations are supplemented here by perfect-gas 
equations-of-state for both the electrons and the ions.  We assume total 
ionization throughout what follows for the sake of simplicity.

\section{The Biermann Effect At Shocks}\label{sec:biermann_at_shocks}

\subsection{Kinematics of the Biermann Effect at Shocks}\label{subsec:kinematics}

By inspection of the Biermann source term in Eq.~(\ref
{eq:Induction_Equation_Biermann}) we see that it is proportional to $\nabla 
n_e\btimes\nabla P_e$. It follows that field generation by the Biermann 
effect can only occur if the gradients of $n_e$ and $P_e$ are not aligned.  
This means that in shocks with planar, cylindrical, or spherical symmetry, 
the field generation rate is zero.  We must therefore treat non-symmetric 
shock surfaces in order to analyze non-trivial cases of field generation.  
Accordingly, in this subsection, we set up the basic kinematics of the 
Biermann effect at general shock surfaces.

\begin{figure}[h]
\begin{center}
\includegraphics[width=0.45\textwidth, bb=60 20 985 725]{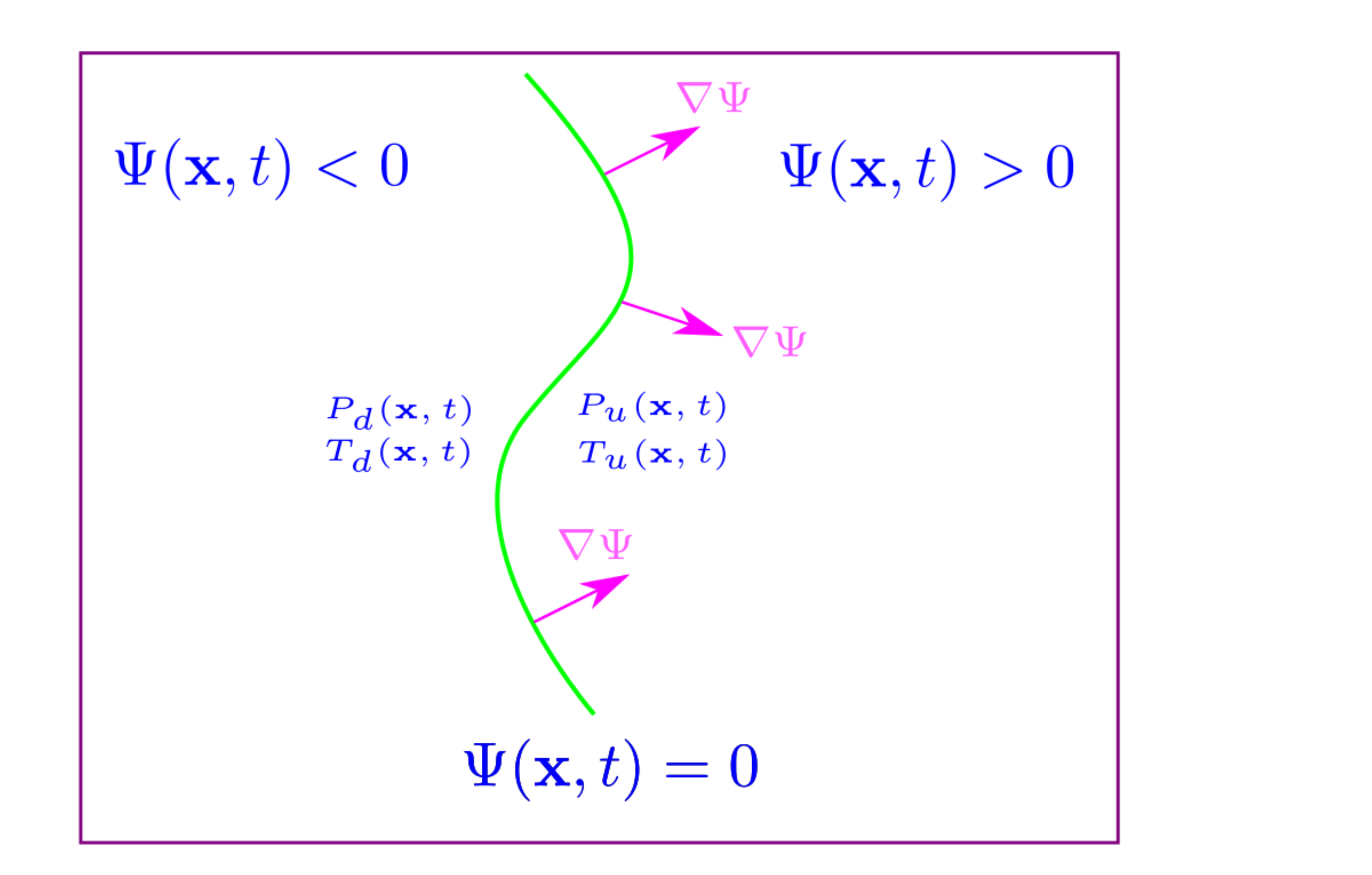}
\end{center}
\caption{Illustration of the kinematics of the Biermann effect at a shock surface.}
\label{figure:Biermann_Kinematics}
\end{figure}

To describe the shock surface, we introduce a level function 
$\Psi(\bx,t)$, and use it to define the shock surface 
$\Psi(\bx,t)=0$.  Note that the function $\Psi$ is of no dynamical 
significance, but is rather simply a mathematical convenience for describing 
the shock surface. We will assume that the shock is moving in the direction 
of the normal vector $\bn\equiv\nabla\Psi/|\nabla\Psi|$, so that the 
region $\Psi(\bx,t)>0$ is upstream, whereas the region 
$\Psi(\bx,t)<0$ is downstream. These definitions are illustrated in 
Figure~\ref{figure:Biermann_Kinematics}.

We denote the local shock speed along $\bn$ by $D$.  By considering
the motion of the level surface $\Psi(\bx,t)=0$ it is not difficult to
show that
\beq
D=-\frac{\partial\Psi}{\partial t}/\left|\nabla\Psi\right|.
\label{eq:Shock_speed}
\eeq

To describe a moving MHD discontinuity that coincides with the surface 
$\Psi(\bx,t)=0$, we will frequently express a field quantity $X$ by the 
decomposition $X=X_u(\bx,t)\Theta(\Psi)+X_d(\bx,t)\Theta(-\Psi)$, where $X_u$ and $X_d$ are
continuous functions, and where $\Theta(\Psi)$ is 
a Heaviside step function. After substituting such expressions into field 
equations, we will find some terms proportional to the Dirac distribution 
$\delta(\Psi)$, resulting from differentiation of the Heaviside functions.  
We will refer to the collected terms of this form as the ``shock part'' of 
the evolution equations.  As will be seen, such terms embody 
Rankine-Hugoniot jumps, which can thus be efficiently extracted for these 
non-symmetric shocks. This trick is also useful for deducing 
hydrodynamic flux terms, as will also be evident below. 

It is convenient to reformulate the Biermann source term using the ideal 
equation of state to replace $n_e$ with $T_e$, the electron temperature.  
Assuming an ideal gas equation of state, the reformulated electric field is
\beq
\bE_B\equiv -(k_B/e)T_e\nabla \ln P_e,
\label{eq:Biermann_E_T}
\eeq 
so that the source term due to the Biermann effect in the induction equation is
\beq
\left.\frac{\partial\bB}{\partial t}\right|_B
=(ck_B/e)\nabla T_e\btimes\nabla\ln P_e.
\label{eq:Biermann_Source}
\eeq
An important reason for this reformulation is that as we saw in \S\ref{sec:plasma_shocks}, 
in the presence of electron
thermal conduction, $T_e$ \textsl{is continuous at the shock}, whereas $n_e$ is
not \citep{Shafranov_Plasma_Shock_1957,zeldovich_raizer}).  As we saw, the continuity of 
$T_e$ is a consequence of the high
mobility of electrons relative to ions, and is therefore part and parcel of the
same approximation that led to the Biermann source term (Eq.~\ref{eq:Biermann_E})
in the first place.  It is central to the developments that follow.

The log electron pressure $\ln P_e$ is discontinuous at the shock, and may be
represented at time $t$ by
\beq
\ln P_e=\ln P_{e,u}\,\Theta\left(\Psi\right) +
\ln P_{e,d}\,\Theta\left(-\Psi\right),
\label{eq:lnPe}
\eeq
where $P_{e,u}(\bx,t)$ and $P_{e,d}(\bx,t)$ are continuous functions.

The discontinuity of $P_e$ leads to a Dirac-$\delta$ singularity in 
$\nabla P_e$.  Using the distributional relation 
$d\Theta(\Psi)/d\Psi=\delta(\Psi)$, we have
\bea
\nabla\ln P_e&=&\ln\left(\frac{P_{e,u}}{P_{e,d}}\right)
\delta\left(\Psi\right)\nabla\Psi\nonumber\\
&&+\Theta\left(\Psi\right)\nabla\ln P_{e,u}
+\Theta\left(-\Psi\right)\nabla\ln P_{e,d}.\nonumber\\
\label{eq:gradPe}
\eea
We may easily interpret the term in Eq.~(\ref{eq:gradPe}) that is proportional
to $\delta(\Psi)$ as the gradient proper to the shock, and the remaining
terms as the gradient in smooth flow.  Since we are interested in
the shock behavior, we define
\bea
\left.\nabla\ln P_e\right|_{Shock}&\equiv&
\ln\left(\frac{P_{e,u}}{P_{e,d}}\right)
\delta\left(\Psi\right)\nabla\Psi\nonumber\\
&=&\ln\left(\frac{\rho_u}{\rho_d}\right)
\delta\left(\Psi\right)\nabla\Psi,
\label{eq:gradPe_Shock}
\eea
where we've used the continuity of $T_e$ and the assumption of complete ionization
to replace the pressure ratio
across the shock with the corresponding compression ratio $\rho_u/\rho_d$.

The electron temperature $T_e$ is continuous across the shock, but its normal
derivative is discontinuous.  We may therefore represent $T_e$ by
\beq
T_e=T_{e,u}\Theta\left(\Psi\right) +
T_{e,d}\Theta\left(-\Psi\right),
\label{eq:Te}
\eeq
where $T_{e,u}(\bx,t)$ and $T_{e,d}(\bx,t)$ are continuous functions, and
where continuity at the shock implies
\beq
\Psi(\bx,t)=0 \Rightarrow T_{e,u}(\bx,t)=T_{e,d}(\bx,t).
\label{eq:Te_continuity}
\eeq

In virtue of the continuity of $T_e$, the gradient $\nabla T_e$ is \textsl{not}
burdened by a Dirac-$\delta$, and we obtain
\beq
\nabla T_e=\Theta\left(\Psi\right)\nabla T_{e,u} +
\Theta\left(-\Psi\right)\nabla T_{e,d}.
\label{eq:gradTe}
\eeq

Inserting Eqs.~(\ref{eq:gradPe_Shock}) and (\ref{eq:gradTe}) into Eq.~(\ref
{eq:Biermann_Source}), we obtain
\bea
\left.\frac{\partial\bB}{\partial t}\right|_{B,Shock}&=&
(ck_B/e)\ln\left(\frac{P_{e,u}}{P_{e,d}}\right)\,\delta\left(\Psi(\bx)\right)
\nonumber\\
&&\times\left[\frac{1}{2}\left[\nabla T_{e,u}+\nabla T_{e,d}\right]\btimes\nabla\Psi\right],
\label{eq:Gen_Rate_1}
\eea
where we've made use of the distributional relation
\bea
\Theta(\Psi)\delta(\Psi)&=&\Theta(\Psi)\frac{d}{d\Psi}\Theta(\Psi)\nonumber\\
&=&\frac{1}{2}\frac{d}{d\Psi}\Theta(\Psi)^2\nonumber\\
&=&\frac{1}{2}\frac{d}{d\Psi}\Theta(\Psi)\nonumber\\
&=&\frac{1}{2}\delta(\Psi),
\label{eq:Theta_times_delta}
\eea
where to get the third line, we've used the fact that the Heaviside function
squares to itself, since its values are either 0 or 1.
This derivation of $\Theta\delta=\delta/2$
is admittedly cavalier in its treatment of distributional quantities, and is presented for brevity.
The result may also be obtained by a limiting procedure, in which the Heaviside
function is represented by the limit of a family of continuous functions -- this
is, after all, the ``Eulerian'' limit of the description of a shock as the viscosity
goes to zero.

Introducing the shock normal vector $\bn$,
and using the fact that the tangential derivative of $T_e$ is continuous (so that
$\bn\btimes\nabla T_{e,u}=\bn\btimes\nabla T_{e,d}$ at the shock), we obtain
\bea
\left.\frac{\partial\bB}{\partial t}\right|_{B,Shock}&=&
\delta\left(\Psi\right)\left|\nabla\Psi\right|\times\nonumber\\
&&(ck_B/e)\ln\left(\frac{P_{e,u}}{P_{e,d}}\right)
\nabla T_e\btimes\bn.
\label{eq:Gen_Rate_2}
\eea

Eq.~(\ref{eq:Gen_Rate_2}) is amenable to a clarifying interpretation, which 
follows from the recognition that if $V$ is a region containing some portion 
$\Sigma$ of the shock surface, and the differential element of surface area 
on $\Sigma$ is $dA$, then 
$\int_Vd^3\bx\,\delta(\Psi(\bx))|\nabla\Psi|=\int_\Sigma dA$, which is to 
say that $\delta(\Psi(\bx))|\nabla\Psi|d^3\bx$ is the differential area 
element on the shock surface.  We may therefore interpret the quantity 
$(ck_B/e)\ln\left(\frac{P_{e,u}}{P_{e,d}}\right) \nabla T_e\btimes\bn$
in Eq.~(\ref{eq:Gen_Rate_2}) as a field generation rate per unit time per 
unit area of the shock surface. This interpretation already allows us to see 
that the Biermann source term should give rise to finite, mathematically 
sensible field generation.

To verify this, we may now obtain the field generation rate due to the passage of the shock 
by inserting the above expressions into the induction equation, Eq.~(\ref
{eq:Induction_Equation_Biermann}), neglecting resistivity. The equation may
be written as
\beq
\frac{\partial\bB}{\partial t}=\nabla\cdot\left(\bB\bu-\bu\bB\right)
+\left.\frac{\partial\bB}{\partial t}\right|_B.
\label{eq:Induction_Redux}
\eeq

In the neighborhood of the shock, we define
\bea
\bB&=&B_n\bn+\bB_{Tu}\Theta(\Psi)+\bB_{Td}\Theta(-\Psi)\label{eq:Shock_B}\\
\bu&=&\bu_u\Theta(\Psi)+\bu_d\Theta(-\Psi)\label{eq:Shock_u}\\
\bu_u&\equiv&\bu_{Tu}+u_{nu}\bn\label{eq:uu_Decomp}\\
\bu_d&\equiv&\bu_{Td}+u_{nd}\bn\label{eq:ud_Decomp},
\eea
where the ``tangential'' components satisfy $\bB_{Tu}\cdot\bn=0$, \break 
$\bB_{Td}\cdot\bn=0$, $\bu_{Tu}\cdot\bn=0$, $\bu_{Td}\cdot\bn=0$. In virtue
of the solenoidal condition $\nabla\cdot\bB=0$, 
$B_n$ is continuous across the shock.  

We transform the advection term $\nabla\cdot\left(\bB\bu-\bu\bB\right)$ in 
Eq.~(\ref{eq:Induction_Redux}) by inserting Eqs.~(\ref{eq:Shock_B}-\ref
{eq:Shock_u}), grouping the Heaviside functions, and collecting the ``shock''
terms proportional to the Dirac-$\delta$ that arises from differentiating the
Heaviside functions.  Using Eq.~(\ref{eq:Shock_speed}) to replace
$\partial\Psi/\partial t$, and using Eq.~(\ref{eq:Gen_Rate_2}), we obtain
\bea
\delta(\Psi)|\nabla\Psi|(-D)(\bB_{Tu}-\bB_{Td})&=&
\delta(\Psi)|\nabla\Psi|\times\nonumber\\
&&\Big\{B_n[\bu_{Tu}-\bu_{Td}]\\
&&-[u_{nu}\bB_{Tu}-u_{nd}\bB_{Td}]\nonumber\\
&&+\frac{ck_B}{e}\ln\left(\frac{P_{e,u}}{P_{e,d}}\right)\nabla T_e\btimes\bn
\Big\},\nonumber\\
\label{eq:Gen_Rate_3}
\eea
whence the Rankine-Hugoniot jump condition, generalized by the Biermann flux is
\beq
\left[(D-u_n)\bB_T\right]^u_d
+B_n\left[\bu_T\right]^u_d
+\frac{ck_B}{e}\nabla T_e\btimes\bn\left[\ln P_e\right]^u_d=0.
\label{eq:Biermann_RH}
\eeq

Here, the notation $[\ldots]^u_d$ means the difference of the upstream and 
downstream values at the shock location. The first two terms in Eq.~(\ref 
{eq:Biermann_RH}) comprise the usual jump condition for the induction 
equation \citep[see Chapter 7 of][for example]{gurnett2005}.  The final term is
the contribution to the jump condition from the Biermann effect, which is seen to 
produce a finite, well-defined discontinuity at the shock.  We may obtain
a useful result by considering the jump in $\bB$ due to a shock advancing into
a quiescent, unmagnetized fluid ($B_n=0$, $B_{Tu}=0$, $\bu_u=0$). The 
downstream magnetic field is then given by
\bea
\bB_{Td}&=&\frac{ck_B}{e}(D-u_{nd})^{-1}\ln\left(\frac{P_{e,u}}{P_{e,d}}\right)
\nabla T_e\btimes\bn\nonumber\\
&=&\frac{ck_B}{eD}\frac{\rho_d}{\rho_u}\ln\left(\frac{\rho_u}{\rho_d}\right)
\nabla T_e\btimes\bn,
\label{eq:Gen_Rate_4}
\eea
where we've used the condition of mass continuity at the shock, 
$\rho_uD=\rho_d(D-u_{nd})$, to replace the term $(D-u_{nd})$, and where 
we've also used the continuity of $T_e$ to replace the pressure ratio in the 
log with the density ratio.

We conclude from the above development that the Biermann source term is 
mathematically well-defined even at weak solution discontinuities, and 
yields definite finite predictions to which a properly designed numerical 
algorithm should be expected to converge.

\subsection{Physical Compatibility of the Biermann Source Term With
Plasma Shock Theory}\label{subsec:validity}

In the Introduction, we raised the question of whether the Biermann source 
term, Eq.~(\ref {eq:Biermann_E}) behaves near a shock according to the 
predictions of kinetic theory, as summarized in \S\ref{sec:plasma_shocks}.  We 
now explain the line of thinking behind this question, and answer it 
definitively.

The induction equation, Eq.~ (\ref{eq:Induction_Equation_Biermann}) can be cast
in conservative form.  In this form, the Biermann source term assumes the form
of the divergence of a flux tensor, whose components are linear in $\nabla P_e$.
It is clear that the construction of these flux components from the gradient of
a discontinuous function is in some way associated with the numerical troubles
that arise with the Biermann effect near shocks.

The key point here is that ``trouble with a flux computed from a derivative'' is actually 
a familiar situation from radiation diffusion, where the radiation flux, which
is proportional to the gradient of the radiation temperature, yields unphysical
(superluminal) fluxes at regions where the temperature changes sharply. 
\citep[see pp. 478-481 of][for example]{mihalas1999}.
Another analogous situation arises with respect
to electron thermal conduction, where the thermal flux from $\nabla T_e$ may yield
unphysically-large transport at shocks, in consequence of the discontinuous change
in $T_e$ \citep[p. 302]{mihalas1999}.
In both of these cases, a straightforward work-around is furnished by a
\textsl{flux limiter}, in which a maximum flux deduced from a kinetic-theory
picture of the transport is used to smoothly cut off the flux in regions
where gradients get large, while leaving the fluxes in smooth regions unaltered.

This is the reason that the question of the validity of the Biermann flux at
shocks is a natural one to ask.  If it were the case that the flux is just wrong
when $|\nabla P_e|$ gets too large, a reasonable approach would be to use
the limiting value for the flux from kinetic theory (Eq.~\ref{eq:EMF}), and impose that flux in
the shock as a cutoff through some kind of flux limiter.
If the electric field due to charge separation given by Eq.~(\ref
{eq:Biermann_E}) results in an EMF across a shock that exceeds this value, 
we could cut it off at this value, and obtain approximate results analogous 
to flux-limited approximation-based results from diffusion-limited radiation 
transport.

Being thus prepared for bad news about the Biermann term from kinetic 
theory, we instead are met by a surprise upon examination of the MHD
flux.  The Biermann electric field at a shock is, according to Eqs.~(\ref
{eq:Biermann_E_T}) and (\ref{eq:gradPe_Shock})
\beq
\bE_B=(k_B/e)T_e
\ln\left(\frac{\rho_d(\bx)}{\rho_u(\bx)}\right)
\delta\left(\Psi(\bx)\right)\nabla\Psi.
\label{eq:E_across_shock}
\eeq
Integrating this over a vanishingly short shock-crossing path $l$ along the 
normal to the shock, we obtain the electromagnetic work done by an electron in
the presence of the Biermann electric field,
\bea
{\cal E}_B&=&e\int_l d\bx\cdot\bE_B\nonumber\\
&=&k_BT_e\ln\left(\frac{\rho_d}{\rho_u}\right),
\label{eq:EMF_Biermann}
\eea
which is the same as the value inferred from kinetic theory, Eq.(\ref{eq:EMF}).

We conclude from this argument that there is no analogy to flux-limited 
diffusion in the behavior of the Biermann source term at shocks.  The 
``plain vanilla'' source gives the correct EMF across the shock, and needs 
no flux limiter to cut it off there.  It should be perfectly possible to 
construct a physically valid algorithm to represent the Biermann MHD source 
term unmoderated by limiters, one that faithfully reproduces the predictions 
of kinetic theory at a shock.

\subsection{Origin of the Numerical Biermann Catastrophe}

As discussed in the Introduction, MHD simulations incorporating the Biermann 
Battery effect have invariably produced catastrophic non-convergent behavior 
as soon as discontinuities develop.  There are two (related) kinds of 
catastrophes that have manifested themselves:  simulations with spherical or 
planar symmetry, in which the charge-separation electric field is 
irrotational and the magnetic field generation rate should therefore be 
zero, develop a non-zero field at shocks with a field intensity that grows 
with increasing resolution \citep{Fatenejad2013}; and simulations of 
High-Energy-Density Physics (HEDP) 
laser experiments containing spatially-inhomogeneous shocks, similar to the 
(highly simplified) simulations presented below in \S\ref
{subsubsec:Ellipsoidal_verification}, in which the field generated at 
shocks, which is not expected to be zero, fails to converge to a finite 
value with increasing resolution -- in order to obtain reasonable results at 
all in such simulations, the cells participating in shocks, contact, and 
material discontinuities must be located at each time step, and the 
generation rate in those cells artificially set to zero \citep 
{Fatenejad2013}.  Such simulation results are clearly not correct, since they
do not treat magnetogeneration due to the ion viscous shock, but at 
least they do converge with resolution.  

Clearly something goes very wrong 
with the numerics when the flow develops discontinuous behavior, whereas the 
behavior of the Biermann term in smooth flow is finite, convergent, and 
stable.
To investigate the origin of these catastrophes, we have addressed the 
following question: suppose the electric field $\bE_B$ of Eq.~(\ref
{eq:Biermann_Source}) is irrotational, in the sense that
\beq
\nabla\btimes\bE_B=\frac{k_B}{e}\nabla T_e\btimes\nabla\ln P_e = 0.
\label{eq:Irrotational}
\eeq
To what extent is this irrotational character preserved after $T_e$ and $P_e$
are separately discretized in a volume-based scheme?

We consider the discretization of two scalar functions $f(\bx)$ and $g(\bx)$ 
that are assumed to have gradients that are everywhere collinear, so that 
there exists some function $\alpha(\bx)$ such that $\nabla g=\alpha\nabla f$.  
A pair of such functions satisfy $\nabla g\btimes\nabla f=0$, so that the 
vector field $g\nabla f$ is irrotational.  The Taylor series of such a pair 
of functions are related to each other by the collinearity constraint. We 
Taylor-expand both functions to third order, average the resulting expansion 
over a grid of control volumes, and compute the ``discrete curl'' -- that is,
the circulation integral about a cell face -- of the discretized ``$g\nabla 
f$'' vector field.  We omit the details of this extremely lengthy calculation
for the sake of brevity, and merely present the conclusions that may be drawn
therefrom.

The result of the calculation is that the leading-order behavior of the 
discrete curl of such a discretized vector field is ${\cal O}(\Delta^2)$, 
where $\Delta$ is the grid spacing.  The coefficient multiplying this term 
is homogeneous of order 3 in the derivatives of $f$ and $g$.  What this 
means is that in regions of space where the two functions are smooth ($C^2$ 
or smoother), the coefficient has a finite approximation in the 
discretization, and the discrete curl tends to zero as $\Delta^2$ with 
increasingly fine resolution, yielding the correct curl of an irrotational
vector field in the limit.

On the other hand, in the presence of a discontinuity, the coefficient of 
the discrete curl has no finite approximation in the discretization, but 
rather diverges as $\Delta^{-3}$ when $\Delta\rightarrow 0$, reflecting the 
meaninglessness of the Taylor approximation in proximity to a discontinuity.
As a consequence, the discrete curl as a whole diverges as $\Delta^{-1}$ with
vanishing $\Delta$ in the neighborhood of a discontinuity.

Setting $f=\ln P_e$, and $g=T_e$, we recognize immediately the source of the 
discrete pathology described above.  The discretization of $P_e$ and $T_e$ 
produces a spurious solenoidal component of $\bE_B$ above and beyond any
real, physically-correct solenoidal component.  This solenoidal anomaly is
small and converges to zero with increasing resolution in smooth flow.  Near
a discontinuity, however, the anomaly is \emph{not} convergent, but rather
diverges as $\Delta^{-1}$.

This is the explanation of the numerical Biermann catastrophe.  It is, 
fundamentally, a completely predictable failure of naive, stencil-based 
approximations to the Biermann source term (Eq.~\ref{eq:Biermann_E}), which are not meaningful when 
Taylor-series approximations to the fluid variables fail in the presence of 
a discontinuity.  The failure is irreducible, and doesn't depend on whether 
the Biermann flux is added to other MHD fluxes or computed separately as a 
split term.

\subsection{The Cure for the Biermann Catastrophe}\label{subsec:biermann_algorithm}

The above diagnosis of the origin of the numerical Biermann catastrophe
does not directly suggest a cure for the problem.  That a cure should exist,
however, is strongly suggested by the fact that the analytic treatment
of the Biermann effect at shocks in \S\ref{subsec:kinematics} leads to finite
and physically-sensible field generation rates.  We therefore now partly retrace 
that development with a view to casting those results in a form suitable for
formulation as a numerical algorithm to be incorporated in a
Godunov-type volume-based MHD scheme.

In order to clarify this proposed strategy, let us briefly review how 
volume-based hydrodynamics/MHD schemes work \citep[for details, see][for 
example]{leveque2002}. Nonlinear systems of hyperbolic PDEs such as MHD are 
prone to develop discontinuities such as shocks and contacts. Direct 
finite-difference discretization approaches break down when this occurs. 
This breakdown is circumvented by appealing to the conservative nature of 
the equations, and replacing the differential equations with their 
corresponding conservation-integral forms, applied to a regular mesh of 
control volumes.  Field quantities are interpreted as volume averages over 
control volumes, and are updated by time-centered (i.e. half-time-step 
forward) fluxes across control-volume faces. These fluxes are finite even 
when solution discontinuities cause the corresponding differential PDE 
source terms to misbehave.

The crucial procedure for carrying out such schemes successfully is the 
computation of the time-centered fluxes by means of the solution of Riemann 
shock tube problems at the cell interfaces.  A Riemann solver takes as its 
input piecewise-constant initial data with a discontinuity at the interface 
and calculates the resulting propagating state, comprising 
piecewise-constant regions separated by discontinuity waves and rarefactions 
belonging to different families of characteristics.  This propagating state 
is used to infer the MHD variables at the interface a half-timestep 
following the initial state. MHD fluxes computed from these variables are 
then used to update the cell-averaged field quantities \citep{leveque2002, 
toro2009}.

Riemann solvers relate adjoining MHD states separated 
by a discontinuity traveling with wave speed $D$ by means of the 
Rankine-Hugoniot relations,
\beq
D\times(\bm{\Phi}_L-\bm{\Phi}_R)=
\bm{F}(\bm{\Phi}_L)-\bm{F}(\bm{\Phi}_R).
\label{eq:RH}
\eeq
Here, $\bm\Phi$ is the state vector of conserved field quantities, 
$\bm\Phi^T\equiv[\rho,\rho\bu^T,\rho{\cal E},\bB^T]$, that is, the mass,
momentum, and energy densities and the magnetic field. $\bm{F}$ denotes the vector
of fluxes corresponding to $\bm{\Phi}$.  The subscripts ``L'' and ``R'' denote
``Left'' and ``Right'' states, respectively, and $D$ is positive for a wave
traveling from Left to Right.\citep{toro2009}.

We can now state a minimal requirement in order to assimilate the Biermann
flux to the other MHD fluxes used to update the fluid state: We need to 
establish the terms added by the Biermann effect to the flux vector $\bm{F}$.
Once these additive fluxes are determined, they can be added to the ideal
MHD fluxes.

The additional flux of $\bB$ due to the Biermann effect is obtainable by 
solving for the shock part of the ``Biermann-only'' induction equation, 
Eq.~(\ref{eq:Biermann_E_T}).  That is to say, we assume the distributional forms 
of Eqs.~(\ref{eq:lnPe}), (\ref{eq:Te}), and (\ref{eq:Shock_B}), plug them into
Eq.~(\ref{eq:Biermann_Source}), and keep only
the terms proportional to the dirac-$\delta$.  The result may be obtained by
inspection of Eq.~(\ref{eq:Biermann_RH}), setting the velocities $\bu$ to zero:
\beq
D(\bB_{Tu}-\bB_{Td})=
-\frac{ck_B}{e}\ln\left(\frac{P_{e,u}}{P_{e,d}}\right)\nabla T_e\btimes\bn,
\label{eq:B_RH}
\eeq
Comparing Eqs.~(\ref{eq:B_RH}) and (\ref{eq:RH}), and 
using the continuity of $\nabla T_e\btimes\bn$, we obtain for the Biermann 
effect flux of $\bB$ along $\bn$
\beq
\bm{f}_\bB(\bn)\equiv -\frac{ck_B}{e}\ln(P_e)\,\nabla T_e\btimes\bn.
\label{eq:B_Flux}
\eeq

This expression is manifestly well-defined at discontinuities.    As in 
Eq.~(\ref{eq:Biermann_RH}), from which Eq.~(\ref{eq:B_Flux}) is derived, the only 
residue of the previously toxic gradient $\nabla P_e$ is contained in the 
benign direction vector $\bn$, which is effectively $-\nabla P_e/|\nabla P_e|$
at the shock. The effect of the cross product $\bn\btimes\nabla T_e$ is to 
eliminate the normal component of $\nabla T_e$, while leaving only the 
tangential components. 
It is clear from the form of 
Eq.~(\ref{eq:B_Flux}) that only the tangential components of $\bB$ are subject
to change according to the Rankine-Hugoniot condition, as expected.

It is convenient for the purposes of algorithmic implementation to re-express
Eq.~(\ref{eq:B_Flux}) in tensor form, as an anti-symmetric tensor $\bm{G}^{(B)}$
whose components $G_{ik}^{(B)}$ express the flux in the coordinate direction
$k$ of the magnetic field component $B_i$:
\begin{equation}
G_{ik}^{(B)}\equiv-\frac{ck_B}{e}\ln(P_e)
\sum_{j=1}^3\epsilon_{ijk}\frac{\partial T_e}{\partial x_j},
\label{eq:B_Flux_Tensor}
\end{equation}
where $\epsilon_{ijk}$ is the totally anti-symmetric Levi-Civita tensor.
$G_{ik}^{(B)}$ is, of course, the spatial part
of the dual Maxwell tensor due to the Biermann effect.
It is worth pointing out that this expression for the flux tensor differs from the 
``naive'' flux 
\begin{equation}
G_{ik}^{(B, Naive)}\equiv+\frac{ck_B}{e}T_e
\sum_{j=1}^3\epsilon_{ijk}\frac{\partial \ln(P_e)}{\partial x_j},
\label{eq:B_Flux_Tensor_Naive}
\end{equation}
obtainable directly from the Biermann electric field
(Eq.~\ref{eq:Biermann_E_T}) by a pure gradient, which has no effect on
the induction equation.  It follows that Eq.~(\ref{eq:B_Flux}) yields a 
general flux of $\bB$ in the direction $\bn$ that can be used anywhere in 
the fluid, not only at discontinuities.

The flux in Eq.~(\ref{eq:B_Flux}) is not the only correction that must be 
applied to $\bm{F}$.  There is also a correction to be applied to the 
\textsl{energy} part of the flux vector, in virtue of the Poynting flux
$(c/4\pi)\bE_B\btimes \bB$ that arises in connection with the 
charge-separation electric field $\bE_B$.  This term is a bit puzzling at
first: since $\bE_B$ is normal to the discontinuity, the Poynting flux is
tangential to the discontinuity, and it is not immediately obvious how such
a flux is to be pressed into service in a Rankine-Hugoniot relation.

Nonetheless, the required flux may be inferred in a manner analogous to the 
calculation by which we derived $\bm{F}_\bB$, above:  We solve the 
``Biermann only'' energy equation
\beq
\frac{\partial(\rho{\cal E})}{\partial t}+
\nabla\cdot\left(\frac{c}{4\pi}\bE_B\btimes\bB\right)=0,
\label{eq:Energy_Biermann_Eq}
\eeq
inserting the distributional forms of Eqs.~(\ref{eq:lnPe}), (\ref{eq:Te}),
(\ref{eq:Shock_B}), as well as
\beq
(\rho{\cal E})(\bx,t)=
(\rho{\cal E})_u\Theta\left(\Psi\right) +
(\rho{\cal E})_d\Theta\left(-\Psi\right).
\label{eq:Energy_Density}
\eeq
After collecting terms proportional to the Dirac $\delta$, we obtain
\bea
D\left[\rho{\cal E}\right]^u_d&=&
\frac{ck_B}{4\pi e}\big[
\left((T_e \nabla\ln P_e)\btimes \bn\right)\cdot\bB\nonumber\\
&&-\left(\nabla(T_e\ln P_e)\btimes\bn\right)\cdot<\bB>\nonumber\\
&&+T_e\ln P_e\bn\cdot\nabla\btimes<\bB>
\big]^u_d,
\label{eq:Energy_RH}
\eea
where in Eq.~(\ref{eq:Energy_RH}) we distinguish between $\bB$ --- the 
magnetic field to the left or right of the interface --- and 
$<\bB>\equiv\frac{1}{2}(\bB_u+\bB_d)$, the average of the upstream and 
downstream fields at the interface.  Comparing Eq.~(\ref{eq:Energy_RH}) and 
Eq.~(\ref{eq:RH}), we obtain for the energy flux along $\bn$ due to the 
Biermann effect
\bea
f_{\rho{\cal E}}(\bn)&\equiv&\frac{ck_B}{4\pi e}\big[
\left((T_e \nabla\ln P_e)\btimes \bn\right)\cdot\bB\nonumber\\
&&-\left(\nabla(T_e\ln P_e)\btimes\bn\right)\cdot<\bB>
\nonumber\\
&&+T_e\ln P_e\,\bn\cdot\nabla\btimes<\bB>
\big].
\label{eq:Energy_Flux2}
\eea

Here we see a potential problem: the third term in Eq.~(\ref 
{eq:Energy_Flux2}) requires an average of the upstream and downstream 
currents $\nabla\btimes\bB$. This is a difficulty if a true flux is to be 
extracted and pressed into service in a numerical scheme, since possible
discontinuities in $\bB$ makes such a term ambiguous.  The resolution of the 
difficulty is that in general, there is always some finite resistivity in 
real plasmas.  As we will see below in \S\ref{subsec:resistive_precursor}, the 
presence of finite resistivity removes the discontinuity in $\bB$, allowing 
us to replace $<\bB>$ with $\bB$. We then obtain for the flux of energy in 
the direction $\bn$
\beq
f_{\rho{\cal E}}(\bn)=\frac{ck_B}{4\pi e}\ln P_e
\left[
\nabla \btimes\left(T_e\bB\right)
\right]\cdot\bn.
\label{eq:Energy_Flux}
\eeq

To summarize, at a cell interface with a normal vector $\bn$, the hydrodynamic fluxes are to 
be adapted to the Biermann effect by adding to the implemented flux vector 
$\bm F$ an additional flux vector $\bm{F}^{(B)}(\bn)$, given by
\newcommand{\strt}{\rule[-2mm]{0mm}{6mm}}
\beq
\bm{F}^{(B)}=
\left(
\begin{array}{c}
\strt F_\rho^{(B)}\\
\strt\bm{F}_{\rho\bm{u}}^{(B)}\\
\strt F_{\rho{\cal E}}^{(B)}\\
\strt\bm{F}_{\bB}^{(B)}
\end{array}
\right)
\equiv
\left(
\begin{array}{c}
\strt 0\\
\strt\bm{0}\\
\strt f_{\rho{\cal E}}(\bn)\\
\strt\bm{f}_\bB(\bn)
\end{array}
\right).
\label{eq:Biermann_Flux_Vector}
\eeq

We expect a discretization of the Biermann effect based on these flux 
expressions to be convergent \textit{so long as the mesh resolves the scales 
on which $T_e$ and $\bB$ are continuous}.  The scale at which $T_e$ is 
continuous is the scale length $\lambda_T$ of the electron thermal 
conduction precursor region \citep 
{Shafranov_Plasma_Shock_1957,jaffrin1964,zeldovich_raizer}, discussed in 
\S\ref{subsec:thermal_structure}.  This length scale is characteristic of the
variation of temperature perpendicular to the shock;  however, it is also the
length scale over which heat diffuses transversely during the time required for
the shock to travel a distance $\lambda_T$ (and hence traverse its own thermal
precursor). It is therefore the scale to be resolved in order for the discretization
of Eq.~(\ref{eq:B_Flux}) to converge.
At coarser resolutions, $T_e$ appears as 
discontinuous at shocks as all the other fluid variables, and the 
discretization discussed here will not yield converged results for $\bB$. 
Only as the thermal precursor zone is resolved can convergence be expected.

In order for the discretization to produce convergent results for energy, it 
is necessary that the simulation resolve the resistive length scale 
discussed in \S\ref{subsec:resistive_precursor}. Note, however, that for 
many physical situations of interest, the magnetic field intensities 
generated by the Biermann effect are so tiny that the correction due to the 
Biermann effect to the energy flux may justifiably be neglected -- it is 
dwarfed by the fluxes of thermal and kinetic energy, and possibly also by 
the advected flux of existing magnetic field energy. Under such circumstances, it 
is an acceptable approximation to neglect the Biermann energy flux, if a 
non-resistive calculation is of interest, or if the resistive scale cannot 
be resolved.

\section{Two Novel Physical Effects Arising From Shock Biermann}
\label{sec:biermann_and_resistivity}

We now exhibit two novel predictions of the theory of the Biermann effect at 
shocks.  They are the existence of two magnetic \textit{precursors} to the 
shock wave, which lead the wave in regions whose extent depends on the 
upstream conditions: a \textit{resistive magnetic precursor}, which arises 
due to magnetogeneration in the shock ``leaking'' into the upstream fluid in 
consequence of the presence of finite non-zero resistivity; and a \textit
{thermal magnetic precursor}, due to smooth-flow Biermann-effect 
magnetogeneration in the fluid motions set up by the electron thermal 
precursor.  We discuss these in turn.

\subsection{The Resistive Magnetic Precursor}\label{subsec:resistive_precursor}

When the resistivity $\eta$ in the induction equation, Eq.~(\ref 
{eq:Induction_Equation_Biermann}) is finite, a new effect appears in the 
vicinity of a discontinuity: a \textit{resistive magnetic precursor} travels 
ahead of the discontinuity, as the impulsively-generated field due to the 
Biermann effect at the shock diffuses out.  This effect is analogous to the 
well-known electron thermal precursor that precedes a plasma shock, which 
was discussed in \S\ref{subsec:thermal_structure}.  The structure of the 
thermal precursor can be estimated by balancing diffusion and advection of 
thermal energy in the frame of the shock, in the presence of impulsive shock 
heating \citep[][pp. 519-520]{zeldovich_raizer}.  Similarly, as we will 
presently see, the structure of the magnetic precursor is a consequence of 
the balance between magnetic field advection and diffusion in the frame of 
the shock, in the presence of impulsive magnetogeneration at the shock.

We consider a small portion of the shock surface, which we will treat as 
approximately planar, and we will work in the rest frame of the shock. We 
assume an approximately steady state in the shock frame, so that we may set 
$\partial/\partial t=0$ in the field equations.  The induction equation then 
becomes
\beq
\nabla\cdot\left(\bB\bu^\prime-\bu^\prime\bB\right)
+\nabla\cdot\left(\frac{c^2\eta}{4\pi}\nabla\bB\right)
+\frac{ck_B}{e}\nabla T_e\btimes\nabla \ln P_e=\bm{0},
\label{eq:Induction_Shock_Frame}
\eeq
where $\bu^\prime$ is the fluid velocity in the rest frame of the shock.

We will keep only the shock part of the Biermann flux in Eq.~(\ref 
{eq:Induction_Shock_Frame}), that is, the right-hand side of Eq.~(\ref 
{eq:Gen_Rate_2}).  In effect, this amounts to assuming that the magnetic 
field generation from the Biermann effect is impulsive at the shock, so that 
the smooth-flow Biermann effect is much smaller than the effect at the 
shock. This approximation is justified if the size $\lambda_T$ of the 
electron thermal precursor is much smaller than the size of $\lambda_B$ of 
the magnetic precursor, so that thermal gradients are unimportant except at 
the discontinuity. The plausibility of this condition will be verified in
\S\ref{subsec:length_scales}.

We may adopt the local simplification $\Psi(\bx,t)=\bn\cdot\bx$, 
$|\nabla\Psi|=|\bn|=1$ for the discontinuity-tracing level function. We 
further adopt coordinates such that $x_1$ is along $\bn$, so that 
$\bn=\bm{e}_1$, and such that $x_2$ is along $\nabla T_e\btimes\bn$.  Away 
from the discontinuity, we ignore all derivatives except for 
$\partial/\partial x_1$.

The presence of the resistive term changes the discontinuous structure of 
the solution. This term is the divergence of the resistive flux of 
$\bB$. That flux behaves analogously to the Fick's law heat flux $-\kappa\nabla 
T$, in that it opposes gradients in $\bB$.  A discontinuous $\bB$ is 
disallowed, because it would result in an infinite restoring resistive flux. 
$\bB$ is therefore now continuous.  By the Rankine-Hugoniot condition for 
transverse momentum \citep[][Chapter 7]{gurnett2005}, it follows that 
$\bu_T^\prime$ is also continuous.  Only $u_n^\prime$ and $P_e$ have 
solution discontinuities in this case. Assuming the upstream fluid is at 
rest in the lab frame, we therefore set $\bu^\prime=u_n^\prime\bm{e}_1$, with
\beq
u_n^\prime=(-D)\,\Theta(x_1)+(u_{nd}-D)\,\Theta(-x_1).
\label{eq:u_decomp_resistive}
\eeq

By the coordinate choice, assuming the far upstream fluid is unmagnetized, 
and in virtue of Eq.~(\ref {eq:Induction_Shock_Frame}), we may set 
$\bB=B\bm{e}_2$.  Putting this all together, we obtain the following structure 
equation for the magnetic precursor:
\bea
\frac{d}{dx_1}\left[\frac{c^2\eta}{4\pi}\frac{dB}{dx_1}
-u_n^\prime B\right] 
+ \delta(x_1)\frac{ck_B}{e} |\nabla T_e\btimes\bn|&&\left[\ln P_e\right]^u_d
\nonumber\\
&&=0.
\label{eq:precursor_structure}
\eea
The meaning of this equation is that the precursor structure is determined 
by the balance of magnetic diffusion and magnetic advection, in the presence 
of impulsive magnetic field generation due to the Biermann effect.

In the upstream ($x_1>0$) region, we may neglect gradients in $\eta$ (which 
depends on $T_e$). Eq.~(\ref{eq:precursor_structure}) becomes
\beq
\frac{c^2\eta}{4\pi}\frac{d^2B}{{dx_1}^2}+D\frac{dB}{dx_1}=0,
\label{eq:precursor_upstream}
\eeq
the solution of which, given $\lim_{x_1\rightarrow+\infty}B(x_1)=0$ 
is
\bea
B_u(x_1)&=&B_0\exp\left(-\frac{4\pi D}{c^2\eta}x_1\right)\nonumber\\
&=&B_0\exp\left(-\frac{x_1}{\lambda_B}\right),\quad x_1>0.
\label{eq:precursor_upstream_solution}
\eea
Here, $B_0$ is an integration constant.  We see that the precursor has an 
exponential shape and a characteristic length $\lambda_B$ given by
\beq
\lambda_B=\frac{c^2\eta}{4\pi D}.
\label{eq:lambda_B}
\eeq
Here, $\eta=\eta(T_{e,+\infty})$ is the value of the resistivity far upstream of 
the discontinuity.

We may obtain a relation for $B_0$ from the jump condition at the 
discontinuity, by integrating Eq.~(\ref{eq:precursor_structure}) across a 
vanishingly-small shock crossing path.  The result is
\beq
\frac{c^2\eta(T_{e})}{4\pi}\left[\frac{dB}{dx_1}\right]^u_d
-\left[u_n\right]^u_dB_0
+\frac{ck_B}{e}|\nabla T_e\btimes\bn|\left[\ln P_e\right]^u_d=0,
\label{eq:precursor_jump}
\eeq
where now $\eta(T_e)$ is evaluated at the shock. This condition together with 
the upstream solution and boundary condition determine $B_0$. 

The ideal MHD jump conditions may be recovered in the limit of 
spatially-constant $\eta$ with $\eta\rightarrow 0$.  In that case, the 
downstream solution satisfying finite-$B$ boundary conditions at 
$x_1\rightarrow -\infty$ is $B_d(x_1)=B_0$, and the first term in 
Eq.~(\ref{eq:precursor_jump}) is just $-DB_0$. Eq.~(\ref{eq:Gen_Rate_4}) 
follows immediately.

\subsection{The Electron Thermal Precursor}\label{subsec:thermal_magnetic_precursor}

One further consequence of the presence of the thermal precursor in $T_e$ 
described in \S\ref{subsec:thermal_structure} is that in general, there will 
be some small amount of magnetic field generated by the smooth-flow Biermann 
effect ahead of the shock, due to plasma motions in the preheated region, 
whose size is $\lambda_T$, given by Eq.~(\ref{eq:Lambda_T}). In general, the 
field intensity due to this effect can be expected to be small compared to 
the field intensity due to the resistive magnetic precursor, since the 
gradients in the thermal magnetic precursor are small compared to those 
available near the shock itself.  In the constant-conductivity simulations 
that we discuss in \S\ref{subsubsec:thermal_magnetic_sim}, we will see that 
the field intensity is in fact quite small.

Nonetheless, this smallness does not necessarily preclude the possibility 
that the thermal magnetic precursor might be experimentally observable.  
Thermal conductivity upstream of a plasma shock is in general not constant, 
but rather depends strongly on electron temperature -- $\kappa_e\sim 
T_e^{5/2}$ \citep{spitzer1962}. The actual structure of the thermal 
precursor is not the gentle exponential decay that one expects for a 
constant $\kappa_e$, but rather exhibits the steep gradient near its outer 
terminus shown in Figure 7.20 of Chapter VII of \citet{zeldovich_raizer}.  
It is possible that this large gradient may be responsible for an 
enhancement of the Biermann effect at the terminus of the thermal conduction 
precursor zone.  This interesting possibility is beyond the scope of this paper.

One consequence of the presence of a thermal magnetic precursor is that even 
in a non-resistive approximation, the notion of an un-magnetized fluid 
upstream of the shock, on which, for example, Eq.~(\ref{eq:Gen_Rate_4}) is 
based, is not strictly correct.  It is a valid approximation only when the 
Biermann generation rate in the thermal precursor is small compared to that 
of the shock.

\subsection{Physical Length Scales}\label{subsec:length_scales}

It is useful to establish some of the relevant length scales under 
conditions of interest. Below, we establish these scales
for conditions prevailing in HEDP experiments.  In \S\ref{sec:biermann_gcf}, 
we will establish the scales characteristic
of galaxy cluster formation.

Using expression for $\eta$ from \citet{huba2007} in the expression for
the resistive magnetic precursor length $\lambda_B$ given
in Eq.~(\ref{eq:lambda_B}), we 
have the following expression for $\lambda_B$ in a  fully-ionized plasma:
\bea
\lambda_B&=&(4/3)(2\pi)^{-1/2}c^2e^2m_e^{1/2}\,Z\ln\Lambda\,D^{-1}(k_bT_e)^{-3/2}
\nonumber\\
&=&16.4\,\mbox{cm}
\times  Z\times\frac{\ln\Lambda}{10}\nonumber\\
&&\times\left(\frac{D}{10^6\,\mbox{cm}\,\mbox{s}^{-1}}\right)^{-1}
\times\left(\frac{k_BT_e}{1\mbox{eV}}\right)^{-3/2},\nonumber\\
\label{eq:lambda_B_numbers}
\eea
where $\Lambda$ is the usual Coulomb logarithm.  Similarly, the expression for 
the electron thermal precursor length scale $\lambda_T$ is \citep[][Chapter 
VII, \S 12]{zeldovich_raizer}
\bea
\lambda_T&=&\frac{2}{5}\frac{\kappa_e(T_{e,0})}{\rho c_{v,e}D}\nonumber\\
&=&\frac{4}{15}\frac{\xi\,(k_BT_e)^{5/2}}{e^4m_e^{1/2}ZA\ln\Lambda 
n_i D}\nonumber\\
&=&5.39\times 10^{-3}\,\mbox{cm}\times\xi\times
Z^{-1}A^{-1}\times\left(\frac{\ln\Lambda}{10}\right)^{-1}\nonumber\\
&&\times\left(\frac{D}{10^6\,\mbox{cm}\,\mbox{s}^{-1}}\right)^{-1}
\left(\frac{k_BT_e}{1\mbox{eV}}\right)^{5/2}
\left(\frac{n_i}{10^{16}\,\mbox{cm}^{-3}}\right)^{-1},\nonumber\\
\label{eq:lambda_T_numbers}
\eea
where $T_{e,0}$ is the electron temperature at the shock, and where $\xi$ is 
a number in the range 1--2.

The numbers in Eqs.~(\ref{eq:lambda_B_numbers}) and (\ref 
{eq:lambda_T_numbers}) have been scaled to conditions that are routinely 
obtainable in large Laser facilities such as Omega and NIF \citep[see, for 
example][]{gregori2012}.  In these conditions, it is clear that the 
assumption $\lambda_T\ll\lambda_B$, required for the validity of the derivation
of $\lambda_B$, is easily satisfied.

It is also immediately apparent that the magnetic precursor length scale is a 
macroscopic scale that can in principle be well-tailored to the physical 
dimensions of the interaction chambers of such facilities.  A 
carefully-designed experiment, which launches a shock into a relatively cold 
upstream plasma (note the dependence of $\lambda_B$ on $T_e$) should in 
principle be able to detect the resistive precursor due to the Biermann 
effect.

Note that as the value of $T_e$ upstream rises, $\lambda_B$ decreases as 
$T_e^{-3/2}$, while $\lambda_T$ \textit{increases} as $T_e^{5/2}$.  We 
therefore only need to increase the upstream temperature to about 7~eV for 
$\lambda_T$ to become comparable to $\lambda_B$, and at warmer temperatures 
than this $\lambda_T$ becomes dominant.  It is therefore plausible
that a physical situation could be created in which the thermal magnetic
precursor might be observable without interference from the resistive magnetic
precursor.

\section{The Biermann Effect And Galaxy Cluster Formation}\label{sec:biermann_gcf}

In the present section, we attempt to estimate the strength of the seed magnetic fields 
that result from a correct treatment of the Biermann effect at shocks in galaxy clusters 
at the time of magnetogenesis ($z \sim 5-3$), and the physical length scales of the
resistive magnetic precursor and the electron thermal precursor in this context.

\subsection{Proto-Galaxy Cluster Field Generation}\label{subsec:PGCFG}

As discussed in the introduction, the Biermann effect has been invoked as the source
of seed magnetic fields that can be amplified by the turbulent dynamo mechanism 
\citep{Kulsrud_1997}.

Given the fact that in these studies, the gradients near shocks that were used 
to calculate the Biermann effect field generation rates are artifacts of the 
hydrodynamic advance schemes, it is possible -- even likely -- that the 
calculated field strengths are incorrect. We now attempt to estimate the 
typical strength of magnetic fields that are to be expected in early galaxy 
cluster formation, based on a corrected treatment of the Biermann effect.  We 
do not perform new simulations, but rather attempt to estimate the typical field strength 
based on published simulation data, in a preliminary effort to determine the 
reliability of field strengths inferred to date in studies of galaxy cluster 
formation.

While this is not an easy task without actual simulation data 
on-hand to analyze, the analysis work described in \citet{Miniati2000}, based 
partly on $\Lambda$CDM simulations described in \citet{Cen_Ostriker_1994} 
provides enough information for us to make a rough estimate of the typical field strength. The 
simulation in question has the properties $H_0=60$~km~s$^{-1}$Mpc$^{-1}$, 
$\Omega_M=0.45$, $\Omega_\Lambda=0.55$, $\Omega_b=0.043$.  We now give a
relatively detailed description of our analysis of the information in
\citet{Miniati2000}, in order to make clear both the basis for the estimated
field strength and the considerable uncertainty that attend these estimates. 

Our starting point is Eq.~(\ref{eq:Gen_Rate_4}), which gives the post-shock 
field strength due to the Biermann effect, assuming an unmagnetized upstream 
fluid. In order to use this equation to estimate field strengths, we need typical
values of the shock speed $D$, the compression ratio $\rho_d/\rho_u$, and the
tangential gradient $|\nabla_\perp (k_BT_e)|\equiv|\nabla (k_BT_e)\btimes\bn|$ that arise
at primordial epochs.

For the sake of exploiting the information available in \citet{Miniati2000}, we 
will choose $z=3$ as our fiducial primordial epoch.  Examining the 
upper-right panel of Figure~8 in \citet{Miniati2000}, we conclude that a 
not-atypical value of $|\nabla_\perp (k_BT_e)|$ is
\beq
|\nabla_\perp (k_BT_e)|\approx\frac{k_B\times 10^6\mbox{K}}{5\mbox{Mpc}}\approx 10^{-35}\,\mbox{erg cm}^{-1},
\label{eq:grad_perp_Te}
\eeq
Neglecting momentarily the fact that the simulation certainly fails to resolve 
the thermal precursor length scale $\lambda_T$ 
(Eq.~\ref{eq:lambda_T_astro_numbers} below) the uncertainty in this estimate 
seems to be a factor of 2 or so.

\citet{Miniati2000} supplies shock speeds at redshift $z=0$ (Figure 5 of 
\citet{Miniati2000}).  These may be approximately shifted to $z=3$ using Figure 
10 of \citet{Miniati2000}, which shows that the kinetic energy processed by 
shocks increased by a factor of 15 between $z=3$ and $z=0$. This suggests that 
temperatures increased by about that much (neglecting shock filling-factor 
differences between the two redshifts), and that velocities increased by about 
a factor of 4.  We had typical shock temperatures of $\sim 10^6$~K at $z=3$ in 
the data leading to the estimate in Eq.~(\ref{eq:grad_perp_Te}). This 
corresponds to a temperature $\sim 10^7$~K at $z=0$. From the top-right 
panel of Figure 5 of \citet{Miniati2000}, this corresponds to a shock speed of 
about $6\times 10^7$~cm~s$^{-1}$ at $z=0$, and hence $D\sim 10^7$~cm~s$^{-1}$ 
at $z=3$.

Putting these numbers in Eq.~(\ref{eq:Gen_Rate_4}), and assuming the strong-shock
limit $\rho_d/\rho_u=4$ appropriate to a $\gamma=5/3$ ideal gas, we obtain
\beq
B=2.9\times 10^{-22}\,\mbox{G}\times\left(\frac{D}{10^7 \mbox{cm s}^{-1}}\right)^{-1}
\times\left(\frac{|\nabla_\perp k_BT_e|}{10^{-35} \mbox{erg cm}^{-1}}\right).
\label{eq:B_estimate}
\eeq

This value should be regarded as a lower limit, 
since, as remarked above, the simulation does not resolve $\lambda_T$.  An 
upper bound on $B$ can be obtained by replacing, in Eq.~(\ref{eq:grad_perp_Te}) 
the 5~Mpc length scale estimated from Figure 8 of \citet{Miniati2000} with 
$\lambda_T$ from Eq.~(\ref{eq:lambda_T_astro_numbers}). This gives 
$B\lesssim 10^{-19}$~G; i.e., a value that is a factor of about $10^3$ higher.  
Obviously, this is a very conservative bound, as it assumes temperature 
fluctuations of order unity over the diffusive scale $\lambda_T$.

It is clear from the very uncertain nature of the factors used to construct 
this estimate that the field strength given in Eq.~(\ref{eq:B_estimate}) is 
itself subject to considerable uncertainty.  A few actual 
MHD simulations with a correct implementation of the Biermann effect seem 
required to establish how much field strength is in fact made available by
the Biermann effect, for turbulent dynamo effects to amplify.  Such simulations 
would be challenging given the spatial scale $\lambda_T$ that needs to be 
resolved.  

\subsubsection{Galaxy Cluster Formation Length Scales}\label{subsubsec:GC_Scales}

We now estimate the physical length scales of the resistive magnetic precursor 
and the electron thermal precursor for the case of galaxy cluster formation using 
the physical parameters inferred in \S\ref{subsec:PGCFG} from \citet{Miniati2000} 
for galaxy cluster formation at $z\simeq 3$. As in \S\ref{subsec:PGCFG} we use 
$T_e=10^6$~K as our reference temperature and $D=10^7$~cm~s$^{-1}$ as our 
reference shock speed. The 
assumed cosmological parameter $\Omega_b=0.043$ corresponds to a proton number 
density $n_i\approx 10^{-7}$~cm$^{-3}$ at $z=0$, and hence to $n_i\approx 
6\times 10^{-6}$ at $z=3$, which we take as the density upstream of the shocks. 
With these parameters, the Coulomb logarithm is $\ln\Lambda\approx 40$. Setting 
$Z=1$, $A=1$, we then have 
\bea
\lambda_B&=&8.20\times 10^{-3}\,\mbox{cm}\times \frac{\ln\Lambda}{40}\nonumber\\
&&\times\left(\frac{D}{10^7\,\mbox{cm}\,\mbox{s}^{-1}}\right)
\times\left(\frac{T_e}{10^6\,\mbox{K}}\right)^{-3/2},
\label{eq:lambda_B_astro_numbers}
\eea
and
\bea
\lambda_T&=&5.01\,\mbox{kpc}\times\xi\times\frac{\ln\Lambda}{40}
\times\left(\frac{D}{10^7\,\mbox{cm}\,\mbox{s}^{-1}}\right)\nonumber\\
&&\times\left(\frac{T_e}{10^6\,\mbox{K}}\right)^{5/2}
\times\left(\frac{n_i}{6\times10^{-6}\,\mbox{cm}^{-3}}\right)^{-1}.
\label{eq:lambda_T_astro_numbers}
\eea

The much more tenuous and considerably hotter primordial plasma reverses the 
relative sizes of $\lambda_B$ and $\lambda_T$ compared to the HEDP case 
considered above: $\lambda_B$ is now negligible, while $\lambda_T$ is an 
astrophysically-significant length. It is worth comparing $\lambda_T$ to 
$\lambda_e$ and $\lambda_i$, the expected Spitzer mean free paths of electrons 
and ions in the plasma, as a consistency check. This is the product of the mean 
collision time $\tau$ with the thermal speed $(3k_BT/m)^{1/2}$.  Using 
the expression for $\tau_e$ from \citet{huba2007}, we have
\bea
\lambda_e&=&\frac{3m_e^{1/2}\left(k_BT_e\right)^{3/2}}{4\sqrt{2\pi}n_e\ln\Lambda e^4}
\times\left(\frac{3k_BT_e}{m_e}\right)^{1/2}\nonumber\\
&=&\frac{3\sqrt{3}}{4\sqrt{2\pi}}\frac{\left(k_BT_e\right)^2}{n_e\ln\Lambda e^4}
\nonumber\\
&=&2.50\times 10^{-1}\,\mbox{kpc}\,\left(\frac{T_e}{10^6\,\mbox{K}}\right)^2
\nonumber\\
&&\times\left(\frac{\ln\Lambda}{40}\right)^{-1}
\times\left(\frac{n_e}{6\times10^{-6}\,\mbox{cm}^{-3}}\right)^{-1}.
\label{eq:MFP}
\eea
The mean free path is independent of particle mass, and for $Z=1$ it is the 
case that $\lambda_e=\lambda_i$. This length is the characteristic size of the 
viscous shock sheath, and is reassuringly shorter than the 
semi-hydrodynamically scaled $\lambda_T$. 

It is also worth considering whether the plasma is collisional, as we have
implicitly assumed by using Spitzer-type transport coefficients.
The magnetic field strength estimated in 
\S\ref{subsec:PGCFG} gives rise to electron gyroradii $\lambda_c$ of 
characteristic size
\bea
\lambda_c&=&\left(\frac{3k_BT_e}{m_e}\right)^{1/2}\left(\frac{eB}{mc}\right)^{-1}
\nonumber\\
&=&41.4\,\mbox{kpc}\,\left(\frac{T_e}{10^6\,\mbox{K}}\right)^{1/2}
\times\left(\frac{B}{3\times 10^{-22}\,\mbox{G}}\right)^{-1}.
\label{eq:gyroradius}
\eea
Since $\lambda_c\sim 100\lambda_e$, the flow is comfortably collisional for the
chosen physical parameters.  An increase in the estimate of $B$ by two orders
of magnitude or more would change this conclusion;  however, it should be noted
that at least at the outset of the process of cosmic magnetogenesis envisioned here,
field strengths were probably small enough that the collisional plasma approach is valid.
Primordial field strengths that may have originated in early-universe phase transitions 
or in inflationary scenarios are poorly constrained by theory and observation, but
are believed to have plausible values $B<10^{-22}$~G in all but a few scenarios
\citep[see][]{Widrow2002,Widrow2012}. It follows that initially, at least, 
the Spitzer-type transport coefficients that we have employed here are likely to be appropriate.
This is in contrast to the apparent suppression of electron conductivity at later
times \citep[see, e.g.][]{markevitch2003,russell2012,sanders2013,zuhone2013}.

\section{Numerical Verification}\label{sec:Numerical_Results}

\subsection{Implementation Using the FLASH Code}

We have implemented the corrected Biermann Effect algorithm, described in
\S\ref{subsec:biermann_algorithm}, within the publically-available FLASH 
hydrodynamic simulation framework \citep{Fryxell_2000,Dubey:2009hh,tzeferacos2014}.  
FLASH is a modular 
and extensible multiphysics scientific simulation software package that has 
been widely used for reactive compressible flows typical of astrophysical 
situations, HEDP applications, cosmology, 
computational fluid dynamics, and fluid--structure interactions. In particular,
FLASH makes available both a 2-temperature single-fluid model and resistive MHD,
which makes it an ideal platform for testing the proposed algorithm.

Below we furnish an outline of the workings of the code, including the 
newly-implemented Biermann effect algorithm.  Much more detail about FLASH is available
in the \textit{FLASH Users 
Guide}\footnote{http://flash.uchicago.edu/site/flashcode/user\_support/}. We plan
to include an implementation of the new Biermann effect algorithm
in a future release of the code.

The 2-temperature MHD model that we use for our numerical tests is expressed 
in conservation form by the following dynamical system:
\begin{eqnarray}
&&\frac{\partial\rho}{\partial t}+\nabla\cdot\left(\rho\bu\right)=0\label{eq:Masscons2}\\
&&\frac{\partial\rho\bu}{\partial t}
+\nabla\cdot\left[\rho\bu\bu-\frac{1}{4\pi}\bB\bB
+\mathbf{1}\left(P+\frac{\bB^2}{8\pi}\right)\right]=0\label{eq:Momcons2}\\
&&\frac{\partial}{\partial t}\left[\rho\left(\frac{1}{2}\bu^2+\epsilon_T\right)+\frac{\bB^2}{8\pi}\right]
+\nabla\cdot\bigg[
\rho\bu\left(\frac{1}{2}\bu^2+\epsilon_T+\frac{P}{\rho}\right)+\nonumber\\
&&\hspace{1.7cm}\frac{1}{4\pi}\left(-\bu\btimes\bB+\frac{c^2\eta}{4\pi}\nabla\btimes\bB\right)\btimes\bB\nonumber\\
&&\hspace{1.7cm}+\frac{ck_B}{4\pi e}\ln P_e\left[\nabla \btimes\left(T_e\bB\right)\right]
\bigg]=0\label{eq:Econs2}\\
&&\frac{\partial\rho s_e}{\partial t}
+\nabla\left(\rho\bu s_e\right)=
-{T_e}^{-1}\nabla\cdot\left(-\kappa_e\nabla T_e\right)\nonumber\\
&&\hspace{3.0cm}+\frac{\rho c_{v,e}}{T_e\tau_{ei}}(T_i-T_e)\label{eq:Secons2}\\
&&\frac{\partial\bB}{\partial t}
+\nabla\cdot\bigg[
\bu\bB-\bB\bu -\frac{c^2\eta}{4\pi}\nabla\bB+\bm{G}^{(B)}
\bigg]=0.\label{eq:Induction2}
\end{eqnarray}
These equations differ from Eqs.~(\ref{eq:Masscons})-(\ref{eq:Induction}) 
only in that the Biermann term in Eq.~(\ref{eq:Econs2}) replaces the one in 
Eq.~(\ref{eq:Econs}) to reflect Eq.~(\ref{eq:Energy_Flux}), while the 
Biermann term in Eq.~(\ref{eq:Induction2}) replaces the one in Eq.~(\ref
{eq:Induction}) to reflect Eq.~(\ref{eq:B_Flux_Tensor}).

FLASH advances the MHD equations using an Unsplit Staggered Mesh (USM) 
scheme that prevents the development of magnetic monopoles due to numerical 
noise in the induction equation, and which is described in \citep{lee2013}. 

The thermal conduction and heat-exchange source terms on the right of Eq.~(\ref{eq:Secons2}) are computed in a 
time-split manner, separately from the ideal MHD advance. In particular, 
thermal conduction advance is fully implicit, and works as described in 
\S17.1.4 of the \textit{FLASH Users Guide}, while heat-exchange operates as
described in \S16.5 of the \textit{FLASH Users Guide}.  The remaining, advective part of 
Eq.~(\ref{eq:Secons2}) is implemented by treating $s_e$ as a passively-advected 
mass scalar.  This equation requires an EOS implementation capable of using 
electron entropy as an input variable. For the current case of perfect-gas EOS 
and total ionization, the standard Sackur-Tetrode equation for entropy is 
implemented in the EOS.

The resistive terms in Eqs.~(\ref{eq:Econs2}) and (\ref {eq:Induction2}) are 
not treated implicitly, but rather are added directly to the MHD fluxes, 
and are thus treated explicitly, 
placing diffusive stability limitations on the maximum timestep.  The 
Biermann terms in Eqs.~(\ref{eq:Econs2}) and (\ref {eq:Induction2}) are also 
added explicitly to the MHD fluxes.  These flux modifications are performed
in such a way as to preserve the solenoidal character of the magnetic field,
by adding them to the ideal MHD Godunov fluxes before these are interpolated
to the edge-centered electric fields of the USM
scheme, after which each face-centered normal magnetic field component is updated
by the circulation integral of the electric field along the edges bounding the
face \citep{lee2013}.

We have not performed an analysis of 
the timestep limitation imposed by the Biermann effect, trusting instead 
that the small magnitude of the effect in the cases considered makes it 
unlikely that Biermann timestep constraints could dominate those due to the 
hyperbolic and diffusive terms.  In this connection, it is noteworthy that the Biermann
term, being quadratic in fluid derivative terms, makes no contribution to wave
dispersion relations obtained by linearization about uniform solutions, and
therefore does not appear to affect ordinary wave speeds.  Intuitively, one
would therefore expect any timestep limitations due to the Biermann effect
to be higher-order corrections, hardly affecting the numerical evolution of
plasmas for which the Biermann term is small in magnitude.  Nonetheless, a full analysis of the effect of the
Biermann term on the hyperbolic structure of the system of PDEs would be 
useful -- possibly even necessary -- for cases of intense field generation.  
Such an analysis would be somewhat complicated by the non-quasilinear structure
of the Biermann term, necessitating an extension of the PDE system to 
quasilinear normal form.

\begin{figure*}[t]
\begin{center}
\begin{minipage}{3.6in}
\includegraphics[width=3.8in,viewport=0 70 1024 800,clip=]{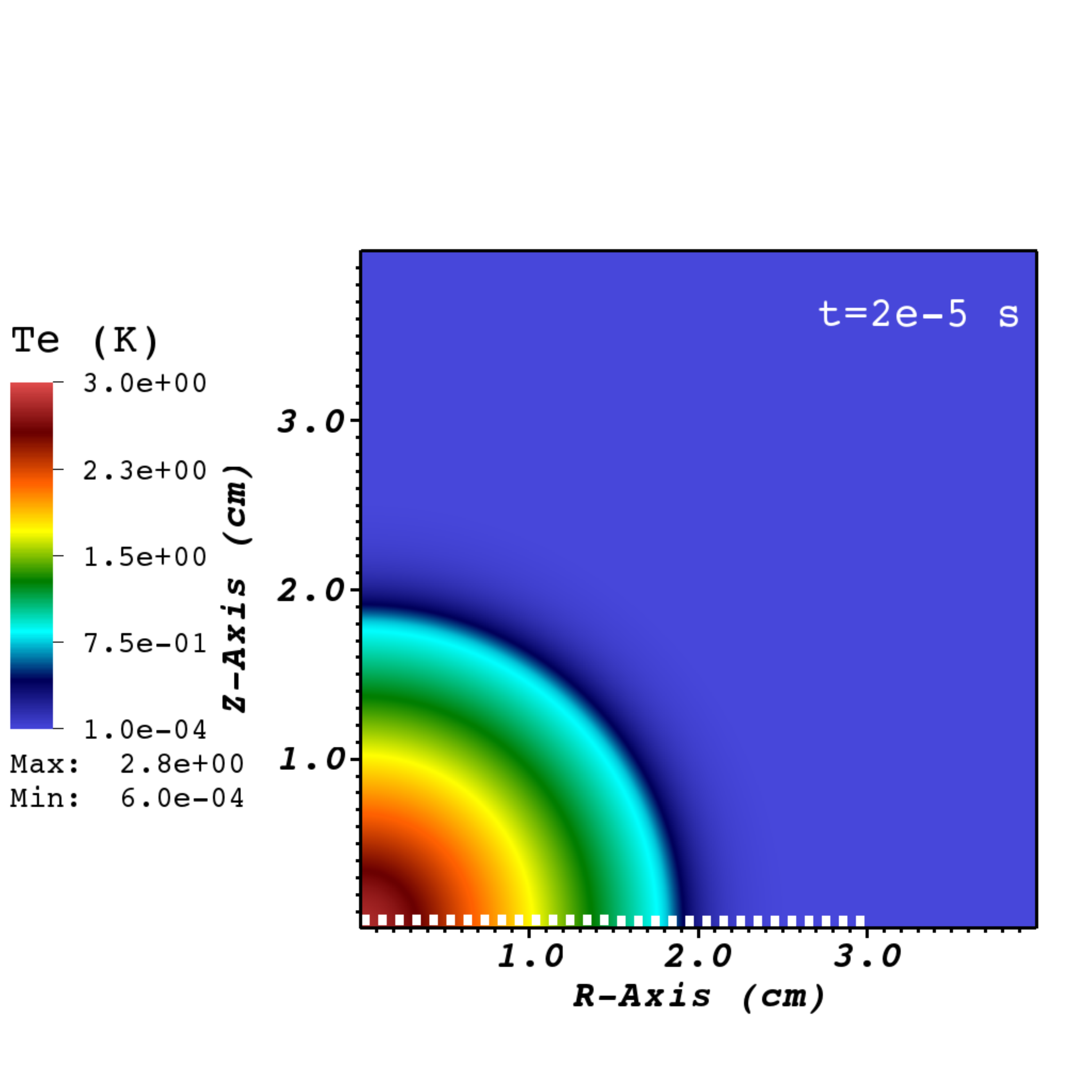}
\end{minipage}
\begin{minipage}{3.0in}
\includegraphics[width=3.0in]{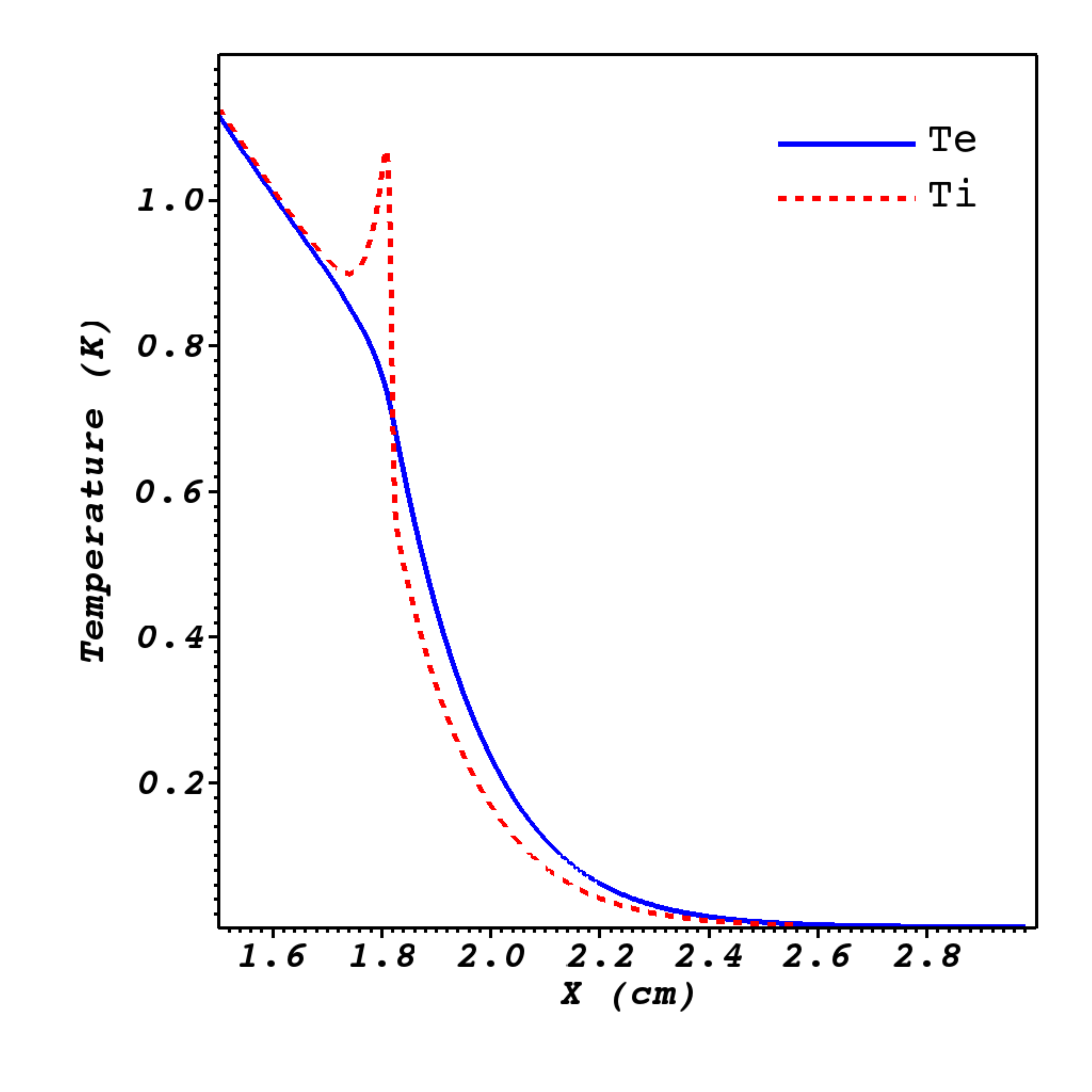}
\end{minipage}
\begin{minipage}[b]{3.6in}
\includegraphics[width=3.8in,viewport=0 70 1024 800,clip=]{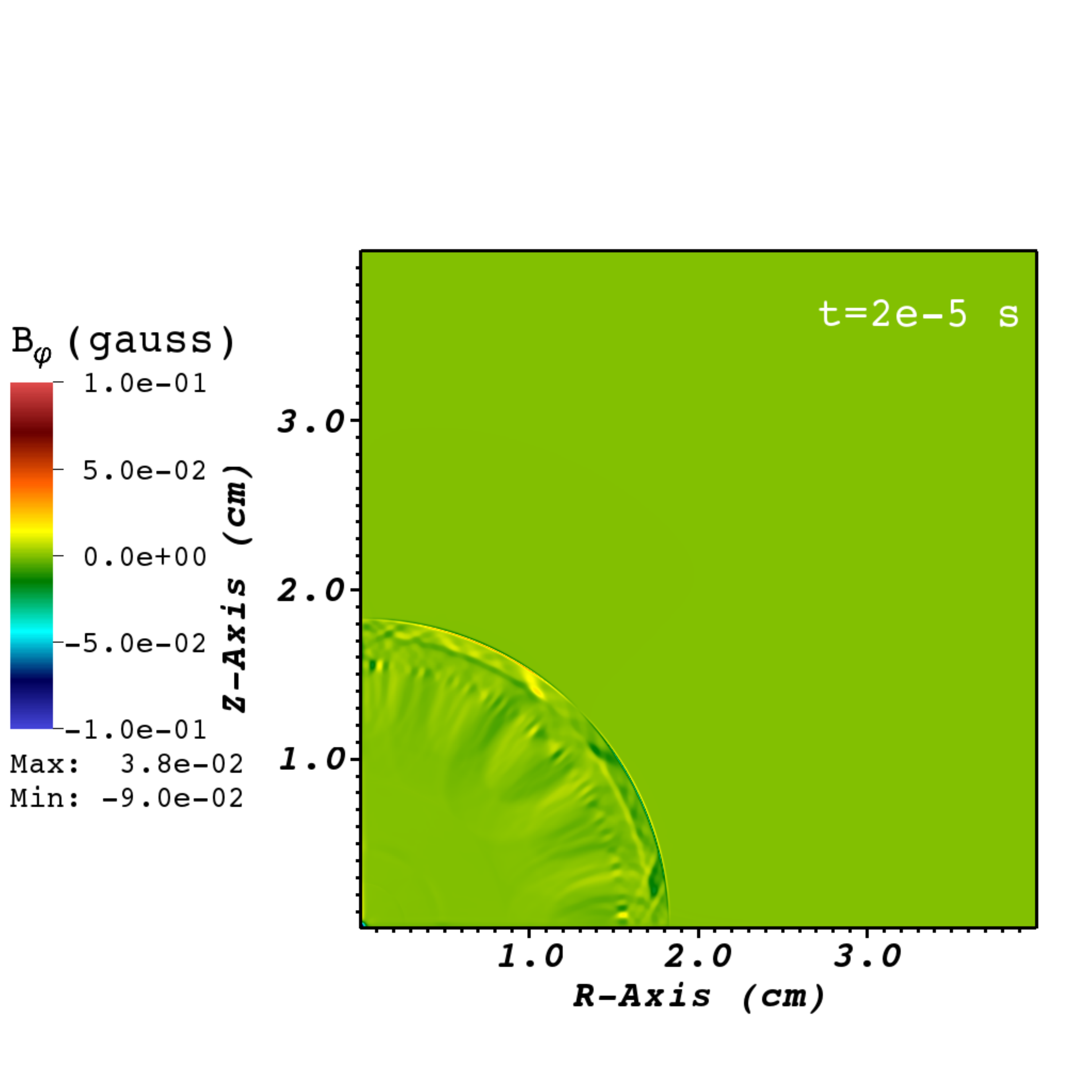}
\end{minipage}
\begin{minipage}[b]{3.0in}
\includegraphics[width=3.0in]{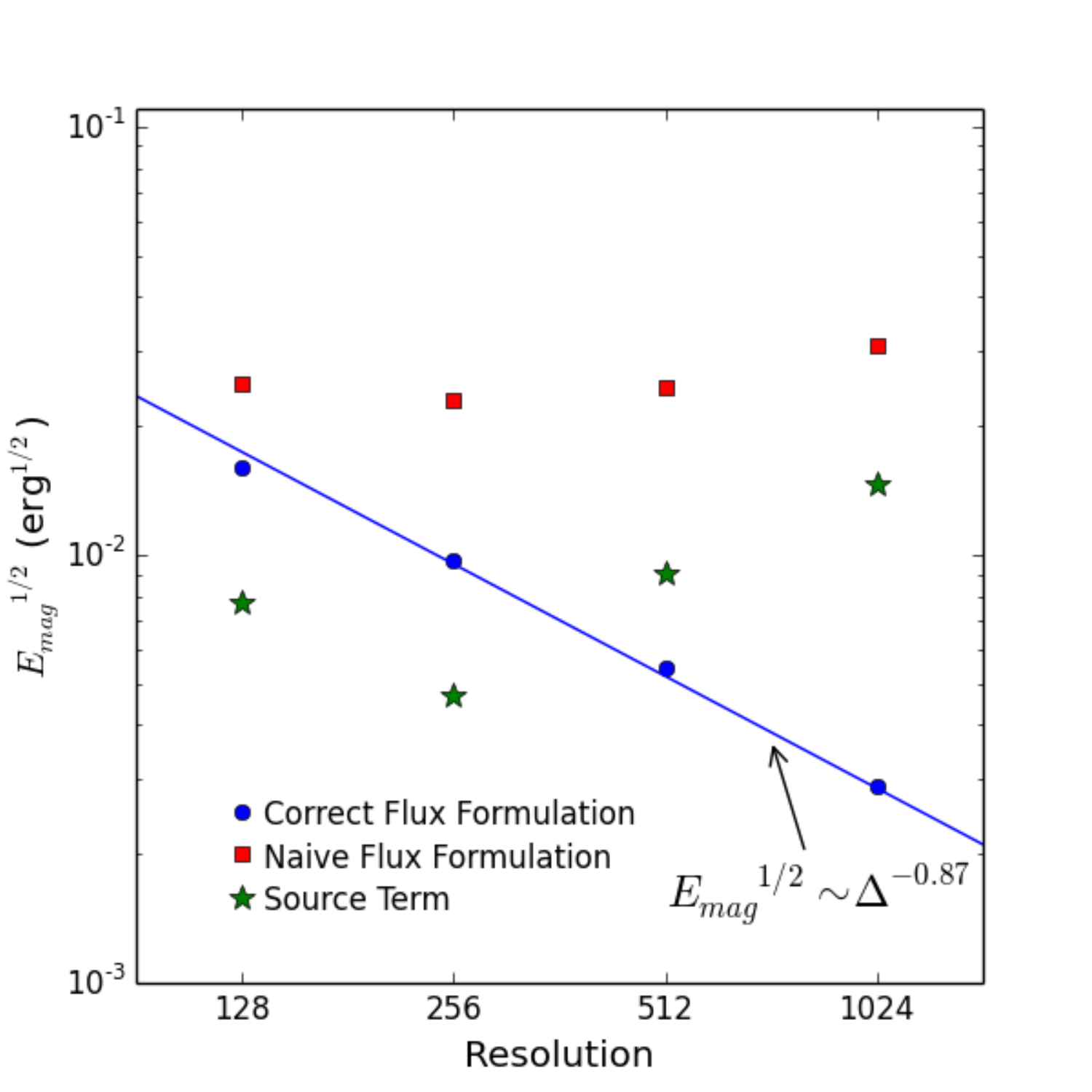}
\end{minipage}
\end{center}
\caption{Two-dimensional simulation of a spherical shock in a 2-temperature Hydrogen plasma.
The images refer to the final timestep, at $t=2\times 10^{-5}$~s, and to the highest
resolution simulation.
Top left: Electron temperature distribution. Top right: Electron and ion temperatures
along the shock-crossing line shown at the bottom of the figure in the top left panel.  Bottom left:  $B_\phi$ due to numerical
noise. Bottom right: Total magnetic energy as a function of simulation resolution.
\label{Figure:Spherical}}
\end{figure*}

\begin{figure*}[t]
\begin{center}
\includegraphics[width=3.5in,viewport=0 70 1024 800,clip=]{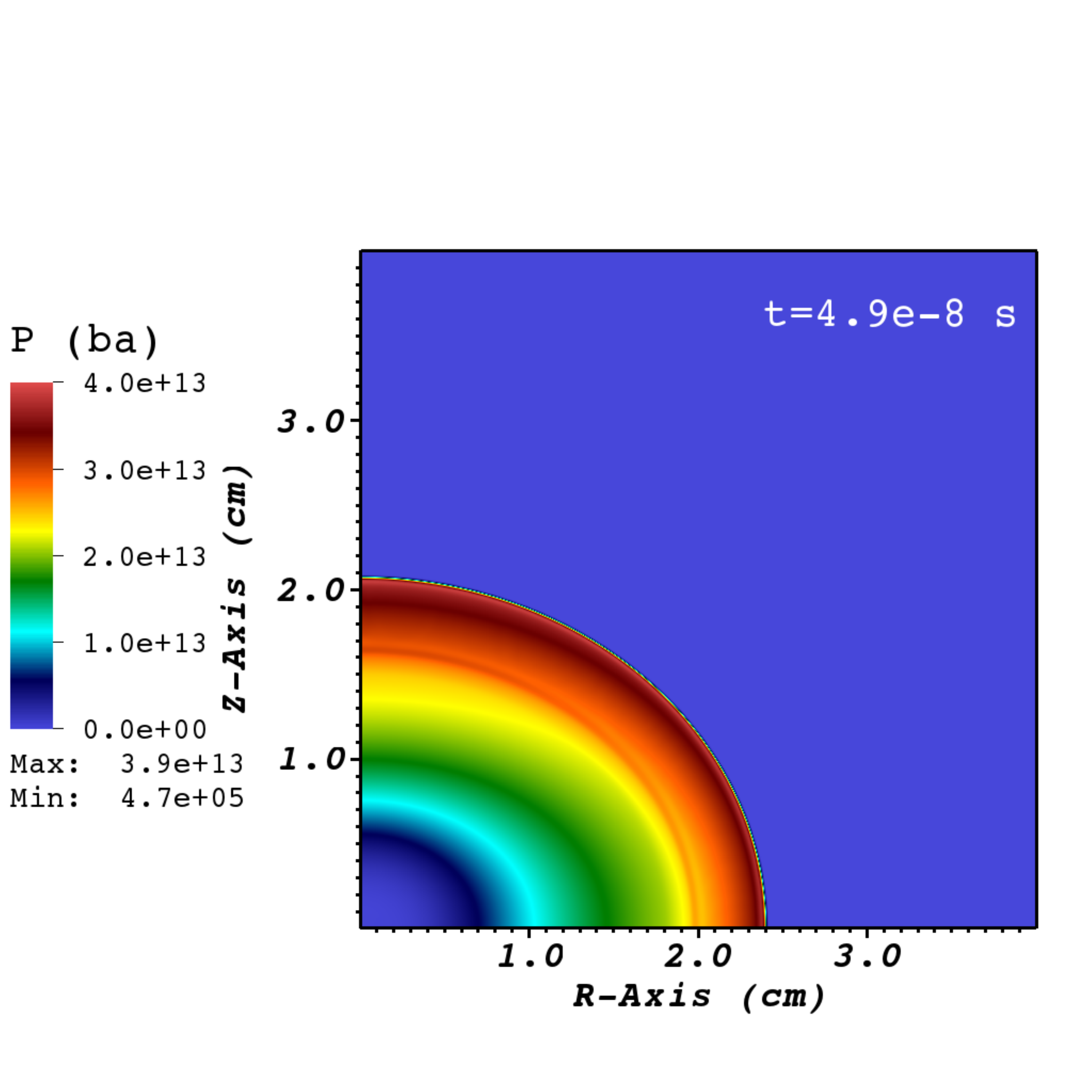}
\includegraphics[width=3.5in,viewport=0 70 1024 800,clip=]{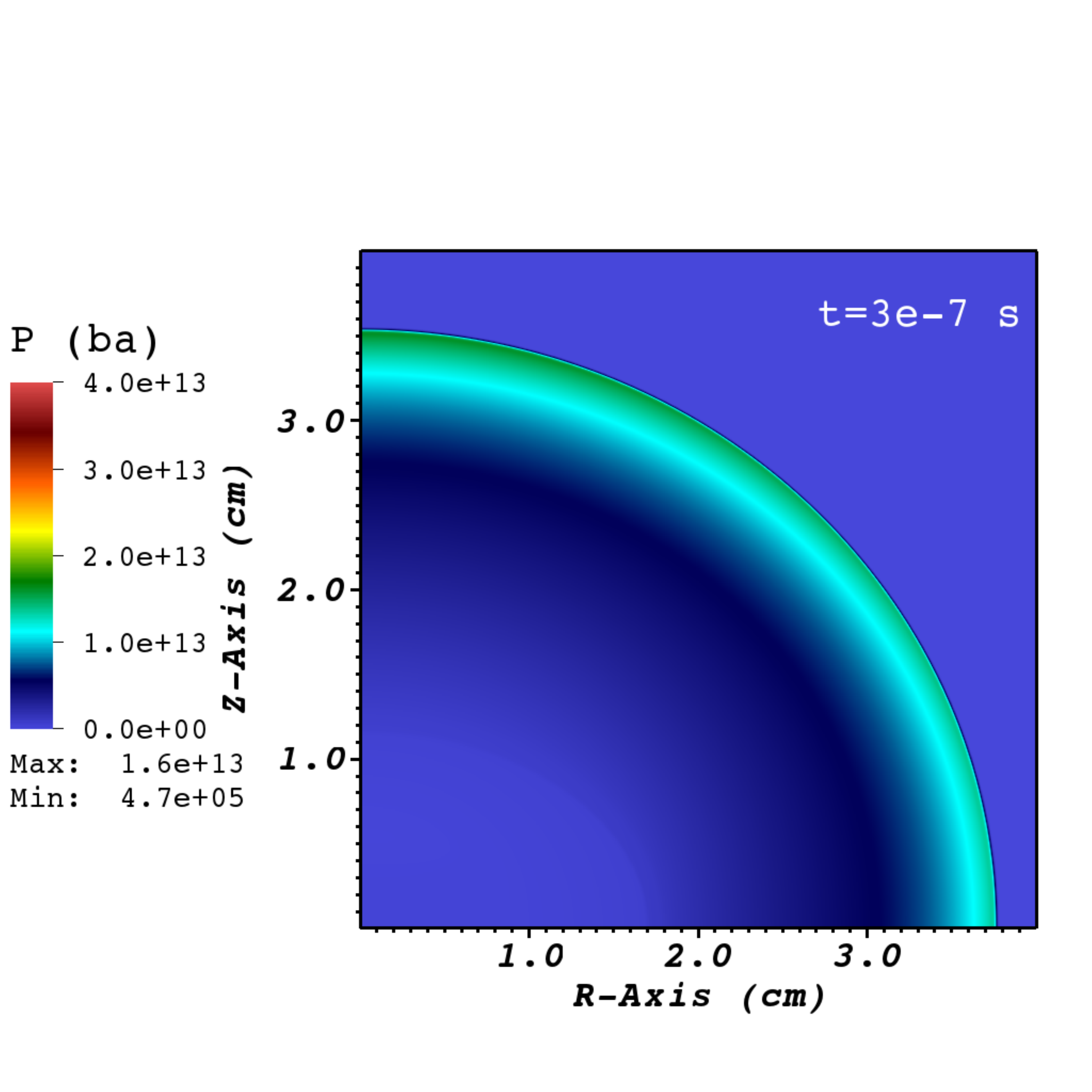}\\
\includegraphics[width=3.5in,viewport=0 70 1024 800,clip=]{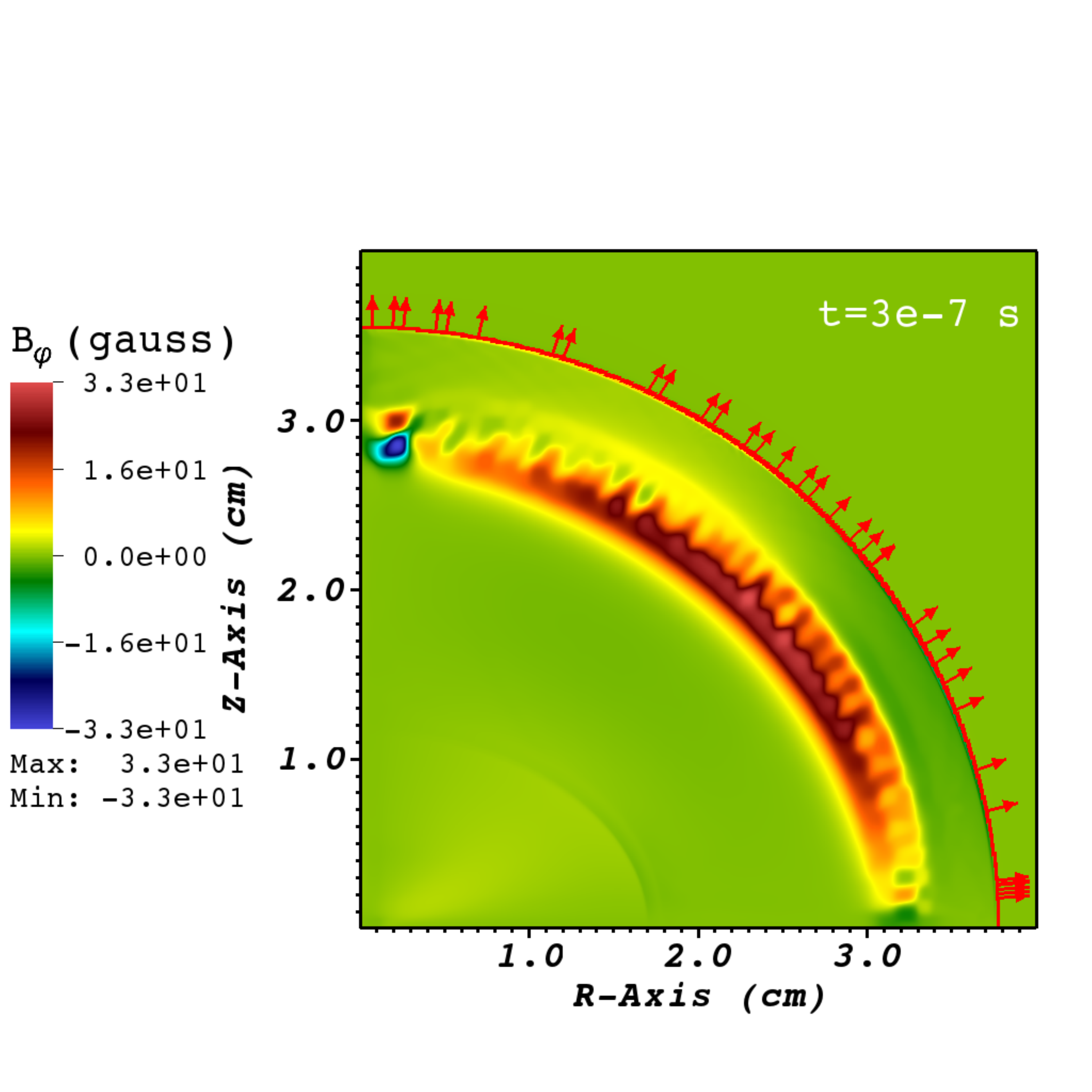}
\includegraphics[width=3.5in,viewport=0 70 1024 800,clip=]{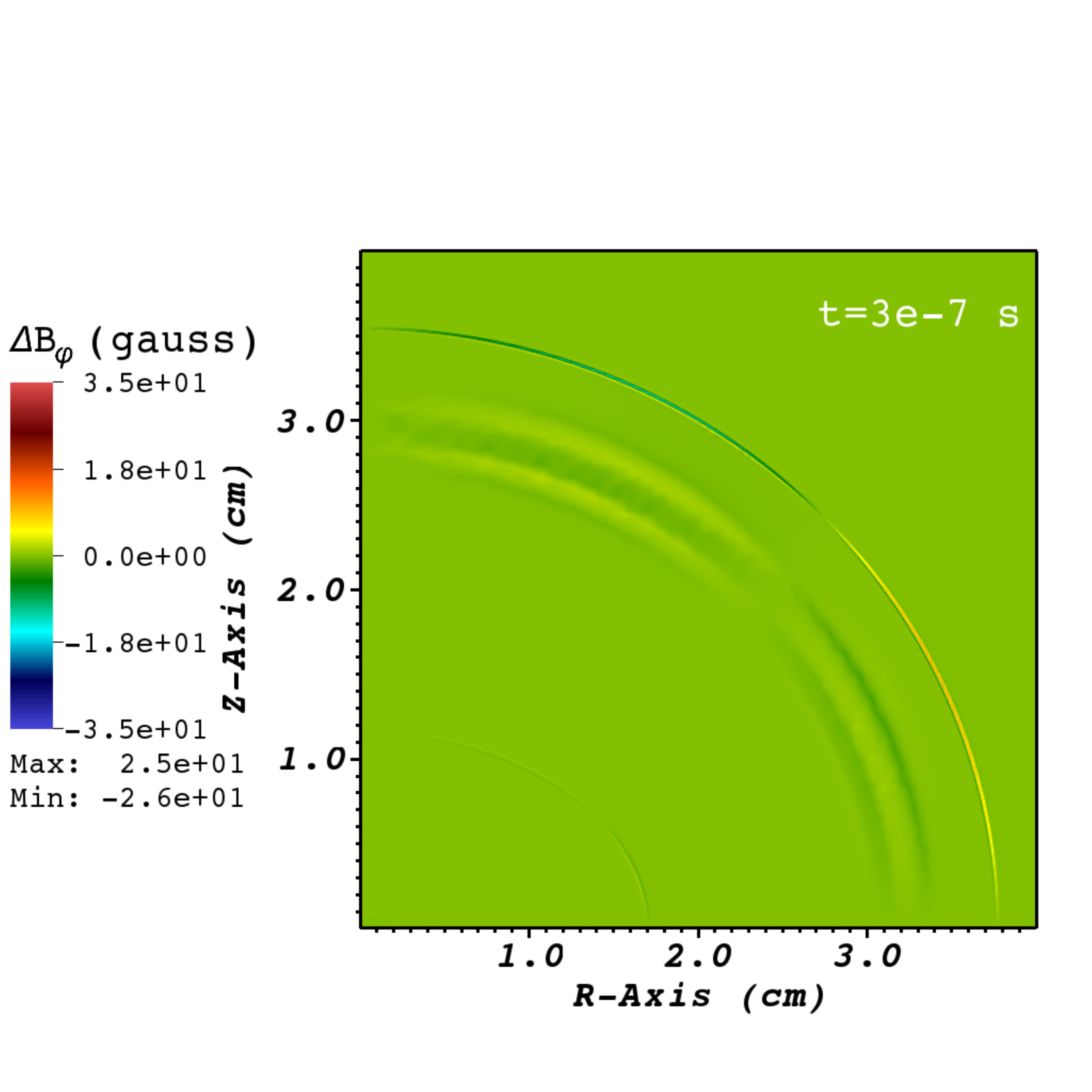}
\end{center}
\caption{Ellipsoidal shock. Top Left: Initial pressure distribution; Top Right:
Final pressure distribution; Bottom Left: Final magnetic field distribution.  The
solid red line displays the location of the shock, while the red arrows display
the direction of the shock normal;
Bottom Right: Difference between final magnetic field distributions computed using
the correct and incorrect Biermann flux.
\label{Figure:NonSpherical_B_Pseudocolor}}
\end{figure*}

\subsection{The Simulations}

The simulations presented below represent idealized situations simplified 
for the sake of verifying the code, and for illustrating the numerical and 
physical principles under discussion. In particular, the conductivity 
$\kappa_e$, electron-ion equilibration timescale $\tau_{ei}$, and the 
resistivity $\eta$ are all treated as constants.  FLASH does have the 
capability to use Spitzer-type functions of the thermal state for these 
parameters, but this would complicate the results unnecessarily without 
adding any real value to these verification tests.

In what follows, we assume fully-ionized Hydrogen -- $A=1$, $Z=1$, adiabatic 
index $\gamma=5/3$, $c_{v,e}=\frac{3}{2}k_BN_A$, where $N_A$ is Avogadro's 
number and $k_B$ is Boltzmann's constant.  Simulations are conducted in 
cylindrical coordinates, assuming azimuthal symmetry, and are therefore 
2-dimensional. In addition, we impose a reflection boundary on the $R$-axis 
(where $R$ is perpendicular distance from axis of rotational symmetry, which 
is to say, $R$ is the cylindrical radius) so that the domain represents a 
hemisphere of a solution with reflection symmetry about that axis.  In every 
case, the domain is 4~cm radius cylinder that extends 4~cm in the $z$ 
direction from the $R$ axis.  The boundary conditions at $R=4$~cm and at
$z=4$~cm are outflow.
 
All of the verification tests that we present are variations on the theme of 
a Sedov-esque explosion:  an 
analytic self-similar Sedov solution is (1) modified to prevent temperatures 
from diverging at the center by setting all variables constant inside some 
chosen radius; (2) smoothed with a Gaussian near the shock, to ease 
instabilities that can otherwise appear near the shock; (3) where required, 
distorted to an ellipsoidal profile capable of producing real Biermann 
effect fields; and (4) scaled to velocities and pressures estimated to 
produce usable simulation times given the resolutions and domain size in 
play. Values of shock ellipticity, $\kappa_e$ and $\eta$ are then chosen to 
produce reasonable thermal and magnetic precursor region sizes, and 
reasonable field generation rates, and values of $\tau_{ei}$ are chosen to 
be harmless.  In every case, we run an initial model for a time without the
Biermann effect, so as to allow transients in the 2-temperature hydrodynamic variables to
subside.  Then we restart the calculation
with Biermann field generation turned on.

Magnetic fields are always azimuthal in these verification tests -- since 
the gradients of $T_e$ and $P_e$ are always in the $R-z$ plane the Biermann
effect only generates non-zero field along the azimuthal direction.

\subsubsection{Null Test: A Spherical Shock}

It is not a simple matter to construct a non-trivial analytic verification 
solution of a plasma generating magnetic field through the Biermann effect. 
A trivial solution, on the other hand, may be straightforwardly constructed 
by imposing symmetry requirements that align the gradients $\nabla P_e$ and 
$\nabla T_e$, thus guaranteeing zero field generation.  We use a spherical
Sedov-like explosion as an example of such a test.  This is in fact the
test that revealed the Biermann catastrophe in the first place \citep{Fatenejad2013}.

We assume an initial shock radius of 1.15~cm, in an ambient medium with 
$\rho=1$~g~cm$^{-3}$ and $P$ set to a negligible value.  We choose an 
initial velocity scale inside the shock that leads to a shock velocity that 
is on average about $3.2\times 10^4$~cm~s$^{-1}$ during the course of the 
simulation.  Other physical parameters are $\kappa_e=1.0\times 10^{12}$~erg~K
$^{-1}$~cm$^{-1}$~s$^{-1}$ and $\tau_{ei}=2\times 10^{-14}$~s. The initial 
solution is advanced for $10^{-5}$~s with no field generation, then 
re-started and run for another $10^{-5}$~s with field generation turned on.  
We repeated this experiment with the correct Biermann flux term of Eq.~(\ref
{eq:B_Flux_Tensor}), with the naive flux of Eq.~(\ref
{eq:B_Flux_Tensor_Naive}), and with the time-split source term of Eq.~(\ref
{eq:Biermann_Source}), so as to compare the performance of the three 
algorithms.

The results of this study are displayed in Figure \ref{Figure:Spherical}. The
top-left panel is a colormap view of the electron temperature $T_e$.  The dashed
shock-crossing line across the bottom of this figure is the location of the data
displayed in the top-right panel, which shows electron and ion temperatures.  The
shock is visible as the peak in $T_i$. The thermal precursor is clearly visible,
as is the continuous behavior of $T_e$, which changes slope at the shock in accordance
with standard plasma shock theory (\citealt[][Chapter VII, \S 12]{zeldovich_raizer}).  
The ion temperature also rises in the precursor region in consequence of the
electron-ion equilibration term.

The lower-left panel of Figure \ref{Figure:Spherical} shows the magnetic 
field strength generated by the correct treatment of Eq.~(\ref 
{eq:B_Flux_Tensor}). There is spurious non-zero field due to numerical noise 
that is clearly being generated.  However, the lower-right panel shows that 
this field is converging to zero with increasing resolution.  This figure 
displays the square-root of the total magnetic energy in the domain.  Since 
the analytic solution for the magnetic field strength is zero everywhere, 
this quantity functions as an un-normalized $L^2$-norm of the difference 
between the analytic and numerical solutions.  The blue circles show the 
convergence with resolution of the correct formulation.  This convergence 
behavior is in contrast to the behavior of the other two algorithms, which 
fail altogether to converge.

\subsubsection{Verification With An Ellipsoidal Shock}\label{subsubsec:Ellipsoidal_verification}      

Next we exhibit simulations designed to produce non-zero values of the 
magnetic field strength.  We start an ellipsoidal shock surface, obtained by 
distorting the spherical Sedov solution.  The initial configuration has a 
semi-major axis (aligned with the $R$-axis) of 2~cm, and a semi-minor axis 
(aligned with the $z$-axis) of 1.667~cm.  The ambient medium is uniform with 
$\rho=1$~g~cm$^{-3}$ and $P$ set to a negligible value.  We choose a 
velocity scale that leads to shock velocities of the about $6\times 10^{6}$
~cm~s$^{-1}$. The higher shock speed is chosen to produce somewhat intense 
magnetic field strengths.  To keep the thermal precursor resolved, we 
increase the conductivity to $\kappa_e=1.0\times 10^{13}$~erg~K$^{-1}$~cm
$^{-1}$~s$^{-1}$.  We also set $\tau_{ei}=10^{-14}$~s.  The initial solution 
is advanced for $4.86\times 10^{-8}$~s with no field generation, then 
re-started and advanced to a simulation time $t=3\times10^{-7}$~s with field 
generation turned on. 

\begin{figure*}[t]
\begin{center}
\includegraphics[width=3.0in]{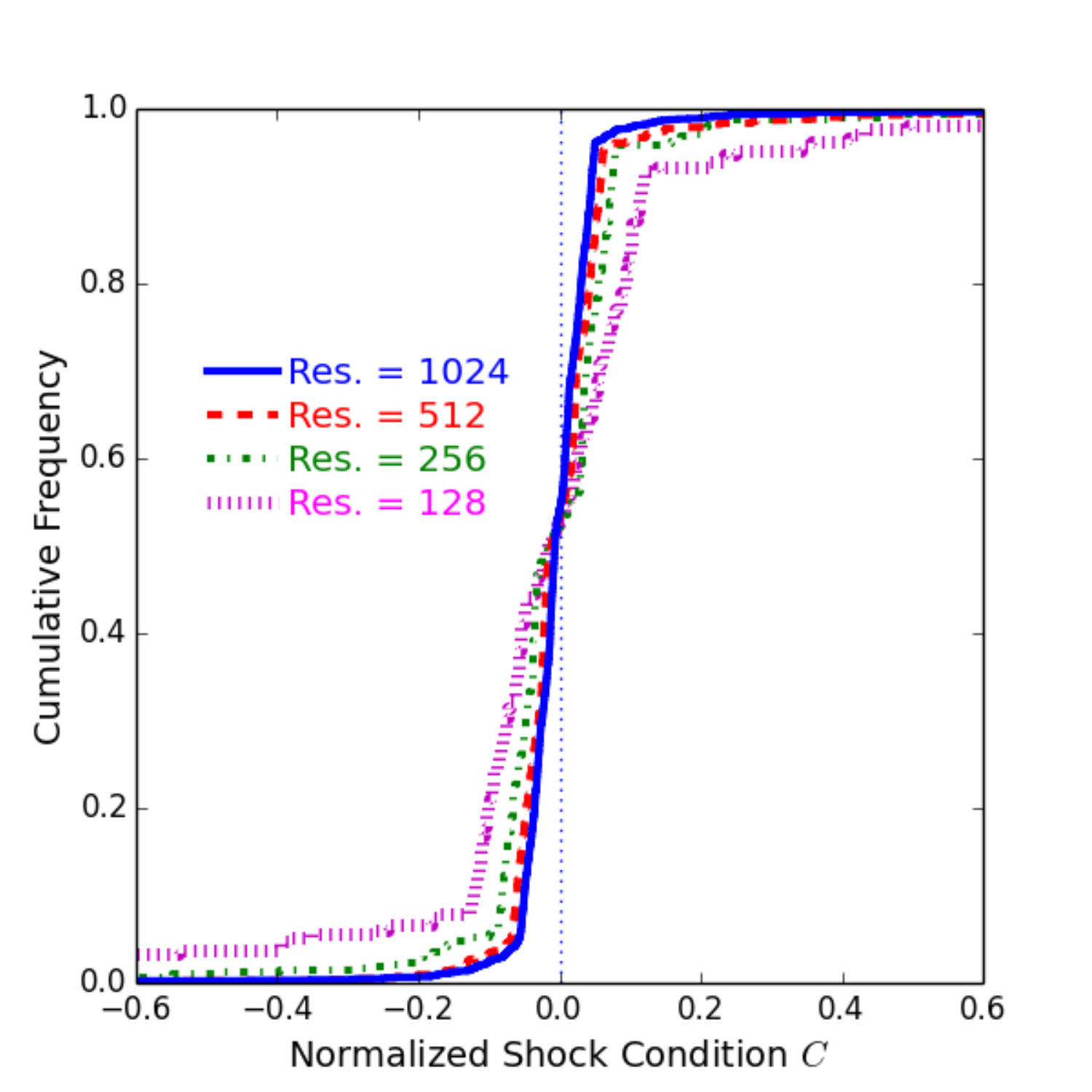}
\includegraphics[width=3.0in]{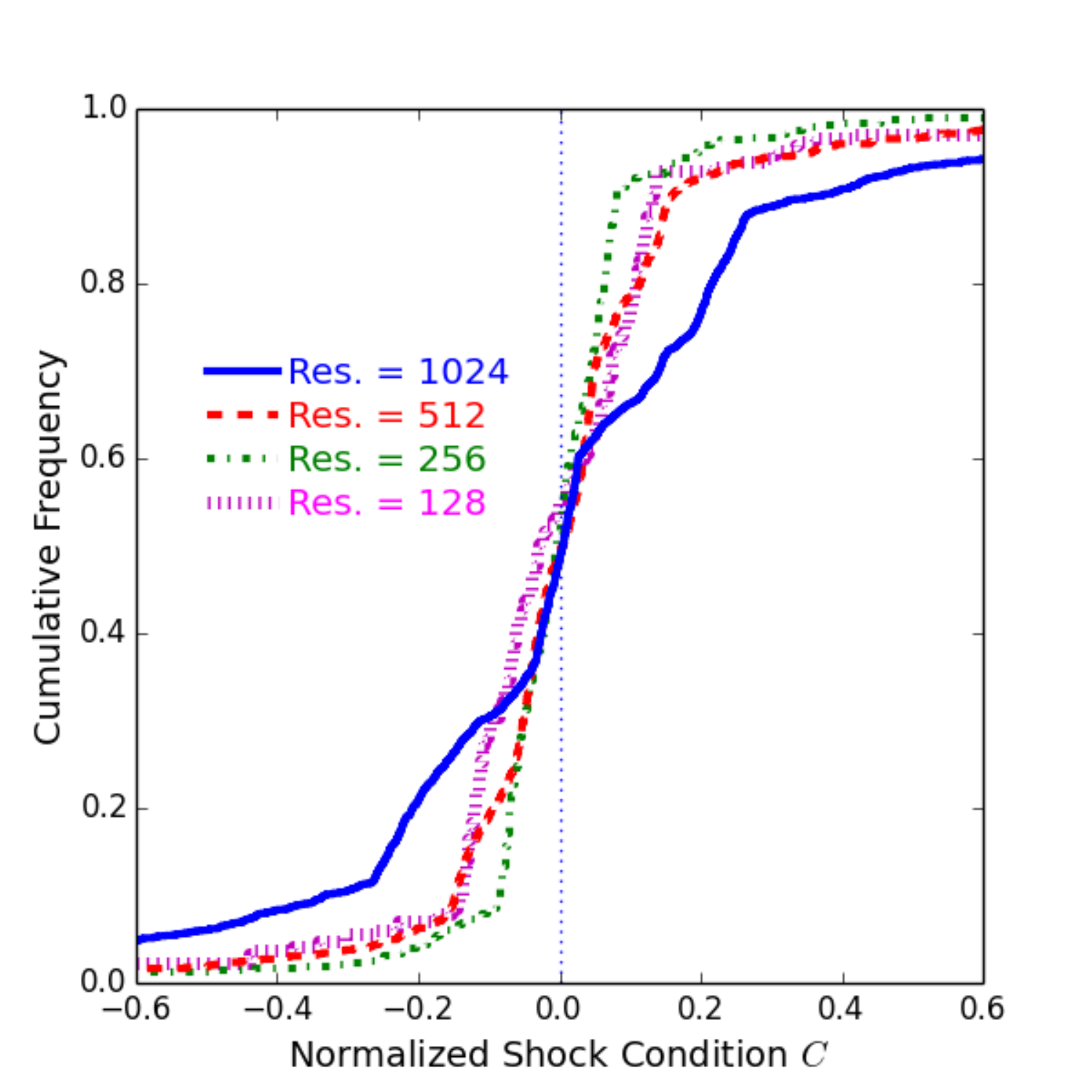}\\
\includegraphics[width=3.0in]{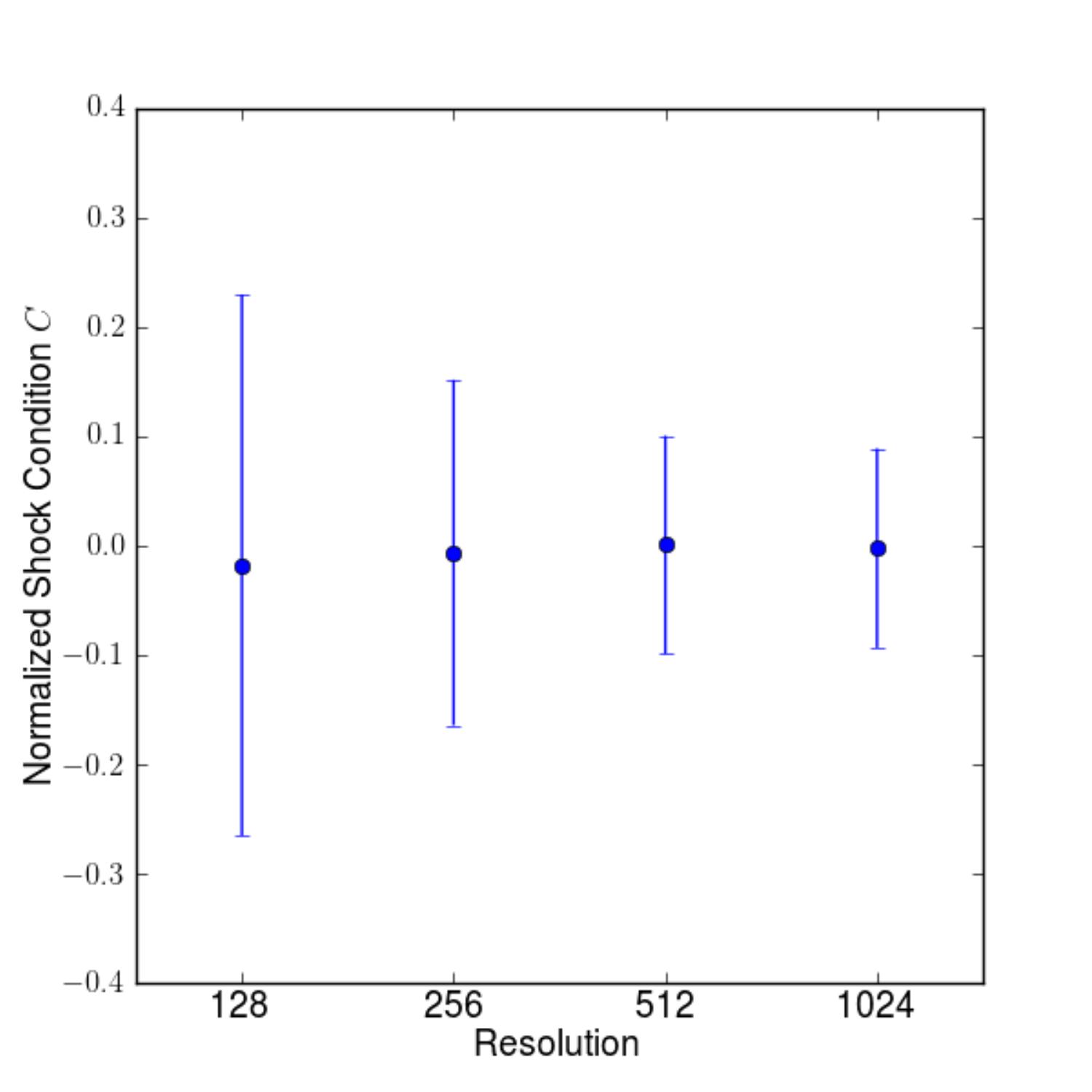}
\includegraphics[width=3.0in]{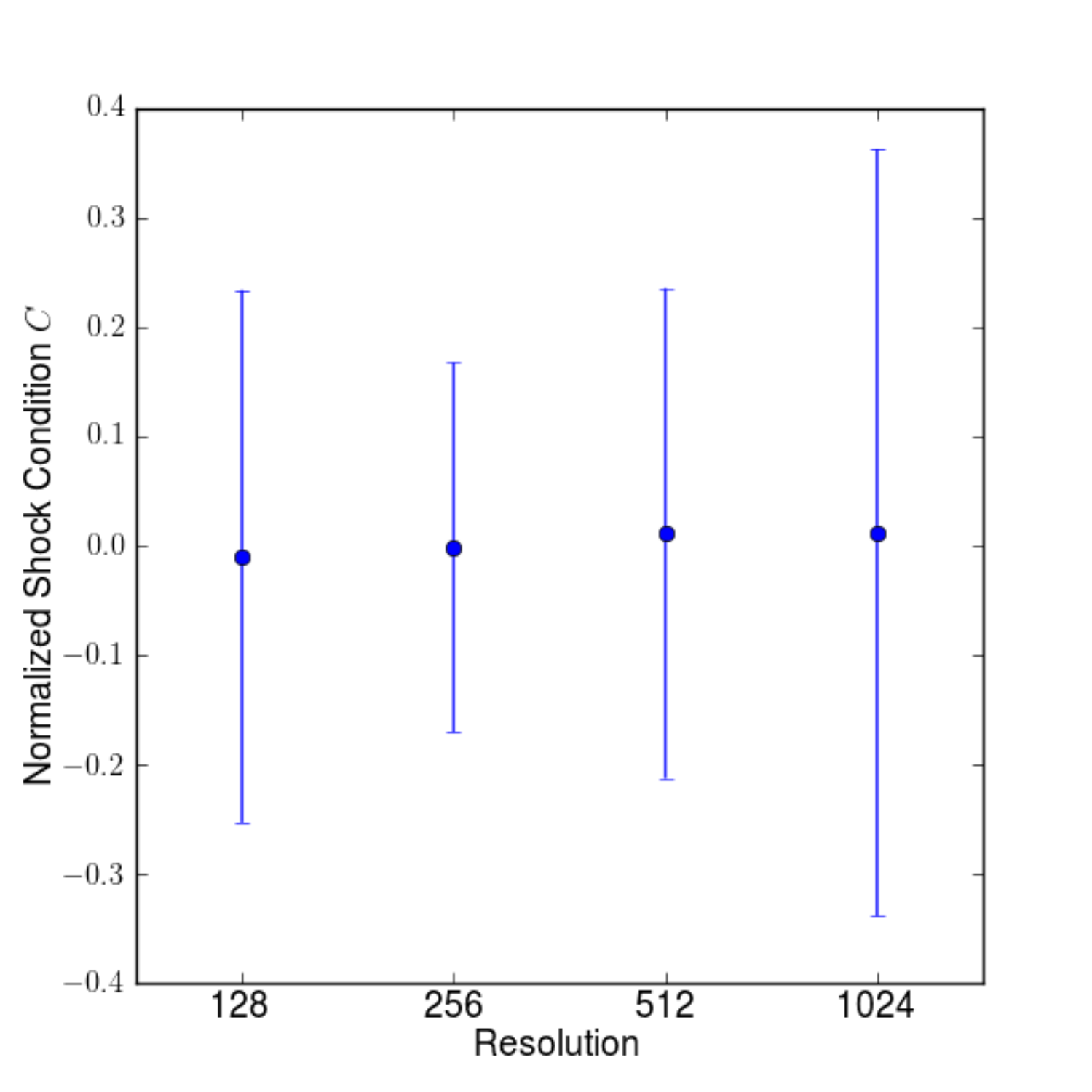}\\
\includegraphics[width=3.0in]{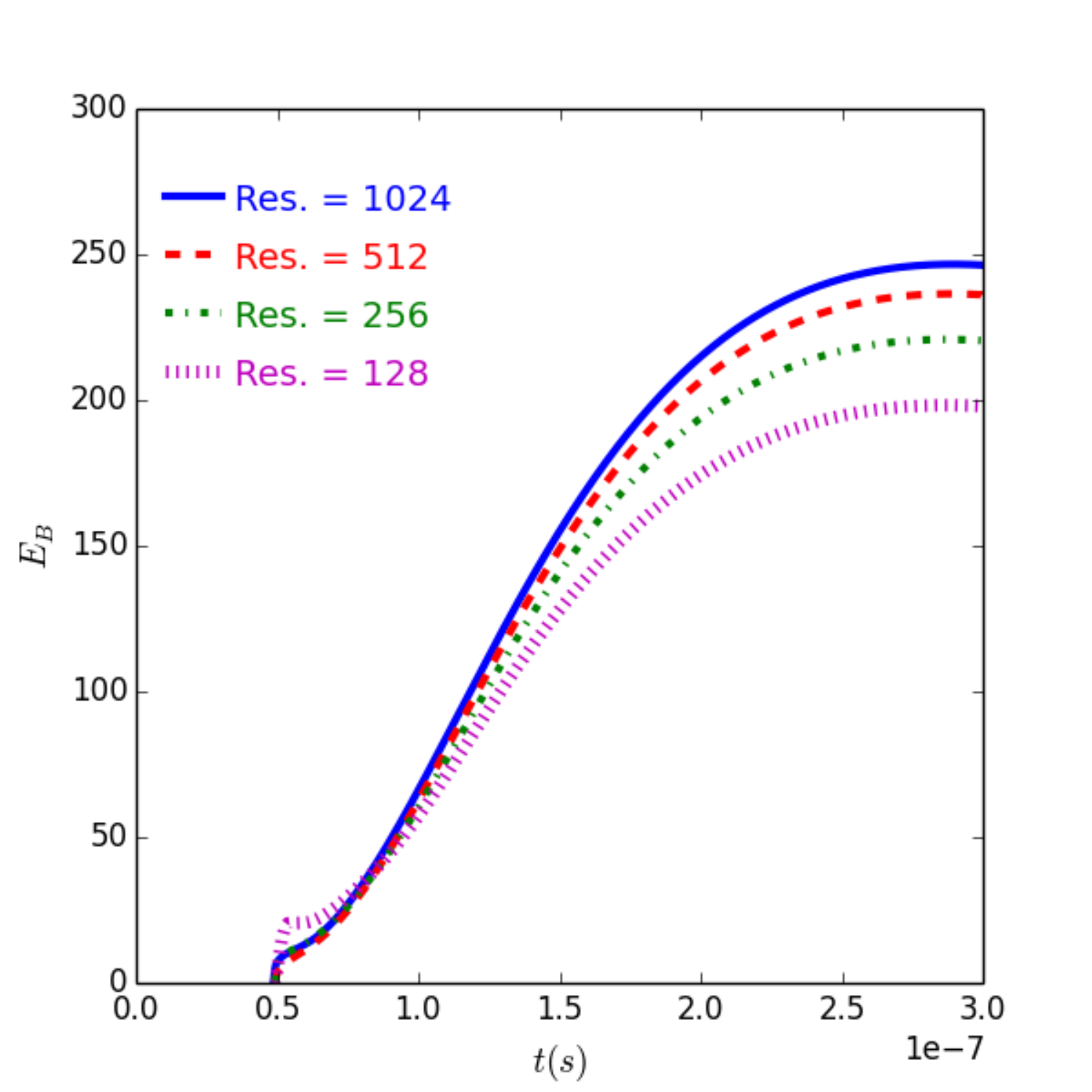}
\includegraphics[width=3.0in]{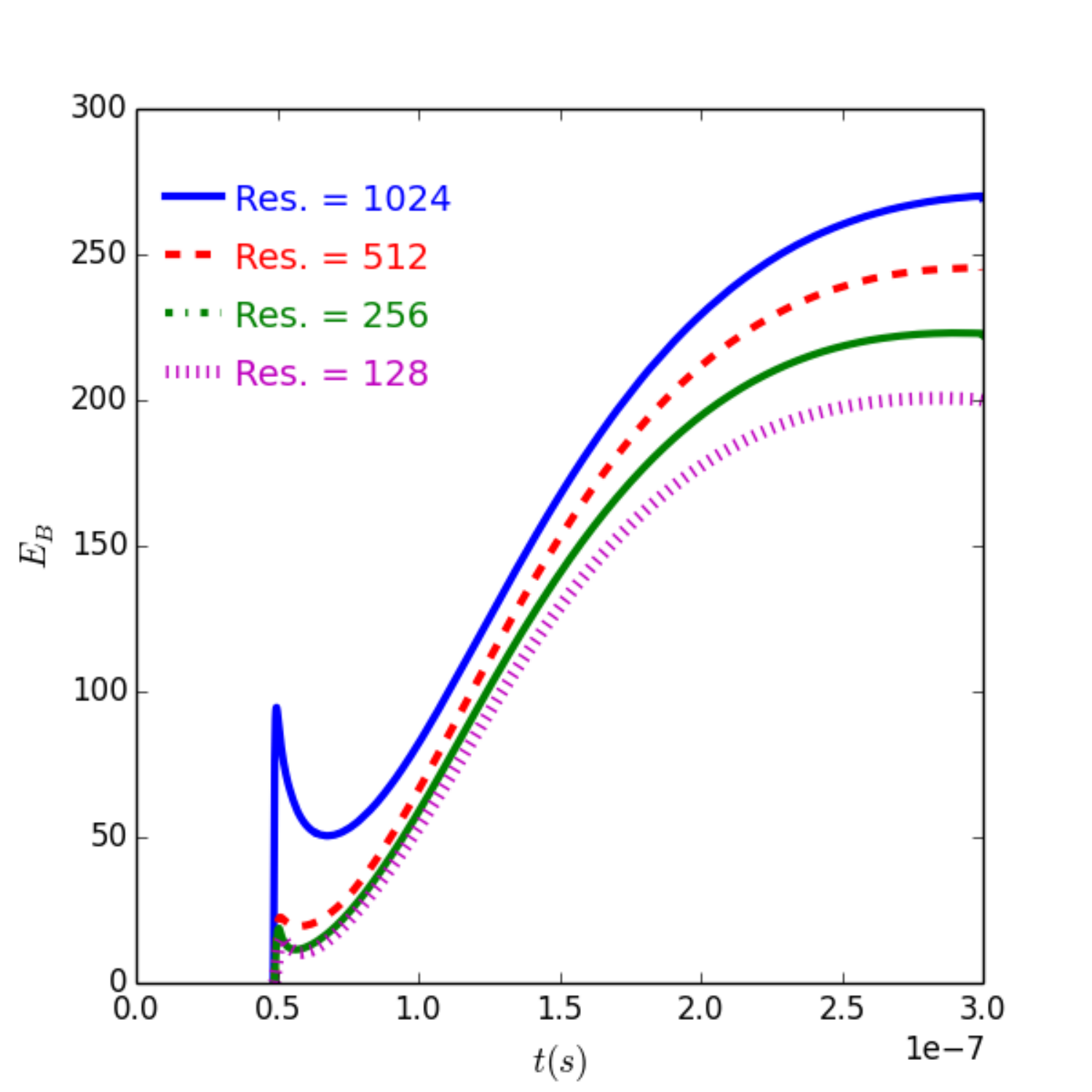}
\end{center}
\caption{Left Panels: Correct Biermann flux term.
Right Panels: ``Naive'' Biermann flux term. Top Panels: 
Cumulative distribution of the normalized magnetic shock condition $C$, defined
by Eq.~(\ref{eq:Shock_Condition_Parameter}), evaluated
at points along the shock surface, for four different resolutions, illustrating
convergence to the correct jump condition. Middle Panels: sample standard deviations of
$C$, as a function of resolution.  Bottom Panels: Total magnetic energy as a function
of simulation time for the four resolutions studied.
\label{Figure:NonSpherical}}
\end{figure*}

\begin{figure*}[t]
\begin{center}
\begin{minipage}[b]{3.4in}
\includegraphics[width=3.55in,viewport=0 70 1024 800,clip=]{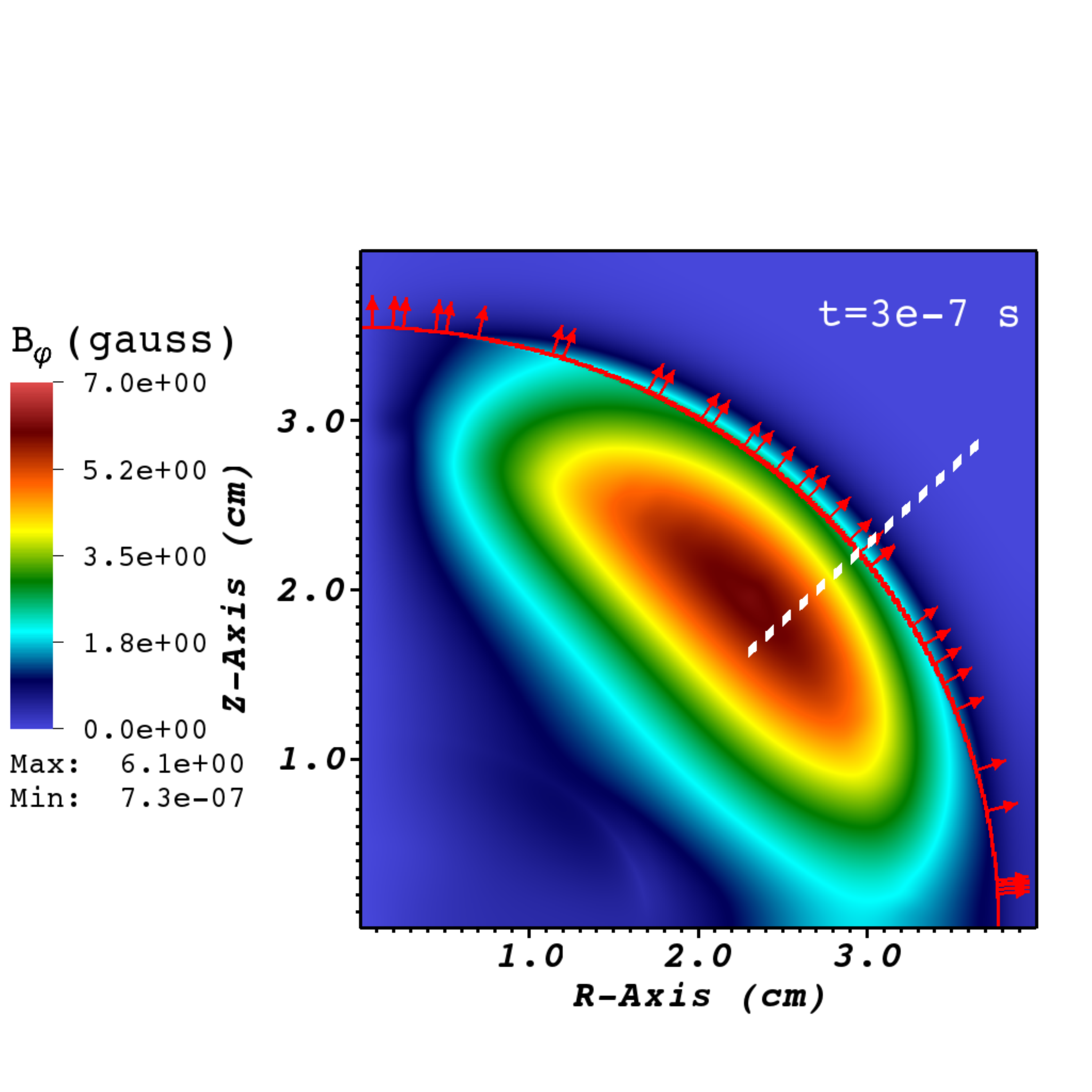}
\end{minipage}
\begin{minipage}[b]{3.25in}
\includegraphics[width=3.0in]{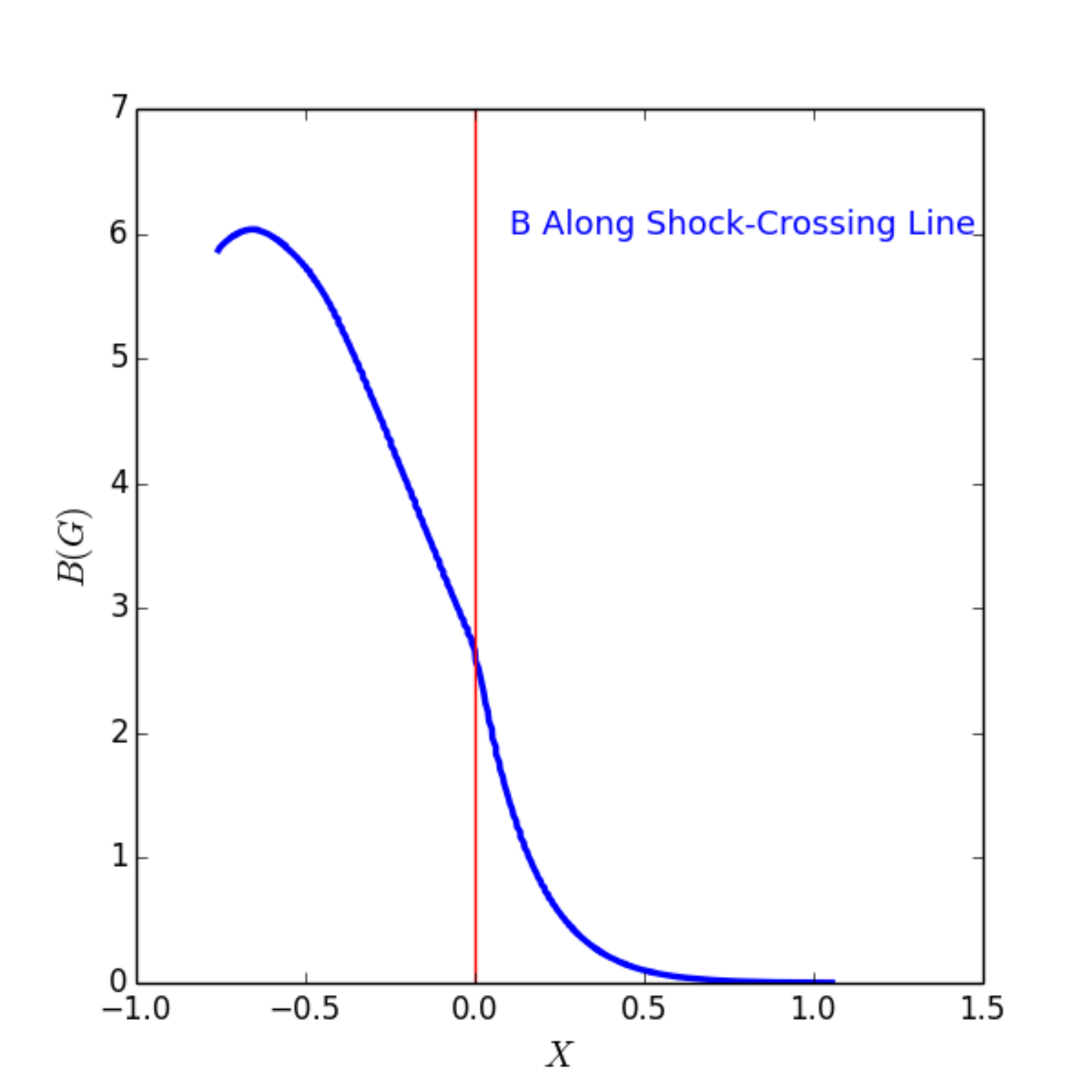}
\end{minipage}
\end{center}
\caption{Left Panel: Magnetic field distribution due to passage of shock, in 
the presence of finite resistivity. The simulation differs from the one 
shown in Figure~\ref{Figure:NonSpherical_B_Pseudocolor} only by the presence of
non-zero resistivity. The solid red line 
shows the location of the shock, while the vectors represent the unit normal 
to the shock.  The magnetic precursor can be clearly seen ahead of the shock 
surface.  The white dashed shock-crossing line illustrates the location of the data 
displayed in the right panel. Right Panel: Magnetic field strength magnitude along the 
white dashed shock-crossing line shown in the left panel.  The position of the shock is 
shown by the vertical solid red line.  The predicted exponential decay of the precursor 
is evident.\label{Figure:Resistive}}
\end{figure*}

The advance of the shock during the period when magnetic field is generated 
is illustrated by the two pressure colormap plots in the top panels of 
Figure~\ref{Figure:NonSpherical_B_Pseudocolor}.  The lower-left panel of 
Figure~\ref{Figure:NonSpherical_B_Pseudocolor} shows the magnetic field 
intensity generated according to the correct flux of Eq.~(\ref{eq:B_Flux_Tensor}), 
ranging into the tens of Gauss.  The solid red line in the figure shows the shock
location, while the red arrows display the shock unit normal vector.  The lower-right panel 
displays the difference between the correct and the naive flux formulations, 
which is substantial at the outgoing shock surface.

We do not have an analytical solution for the magnetic field distribution to 
compare to the output of these simulations.  We do, however, have the 
relation between shock quantities expressed by Eq.~(\ref{eq:Biermann_RH}), which 
determines the jump condition for $\bB$.  By locating points adjoining the 
shock, and computing the local shock velocity at those points, we can then 
verify that Eq.~(\ref{eq:Biermann_RH}) in fact obtains, to some accuracy 
limited by the numerical approximation.  This is in principle a non-trivial 
verification, since the code does not know about the jump condition Eq.~(\ref
{eq:Biermann_RH}) as such -- it only knows the fluxes expressed by Eq.~(\ref
{eq:B_Flux_Tensor}), which were inferred from the jump condition. Recovery 
of the jump condition thus constitutes a verification that the code 
correctly models the theory.

To perform this verification, we first locate cells adjoining the shock by 
the method described in Appendix \ref{appendix:Shock_Detection}, which 
yields a 1-cell wide shock surface by fitting the mass, momentum, and energy 
Rankine-Hugoniot conditions to the neighboring data, using a speed-of-sound 
weighted inner-product on the state space, and treating the shock speed $D$ 
as a free fit parameter.  The shocked cells are the ones that fit the R-H 
conditions with small fit residuals ($R^2<0.01$), together with some other natural
auxiliary consistency conditions described in the Appendix.  The fitted shock speed
is then used in the verification of Eq.~(\ref{eq:Biermann_RH}).

In the absence of normal component field $B_n$, Eq.~(\ref{eq:Biermann_RH}) becomes
\bea
&&(D-u_d)B_d - (D-u_u)B_u\nonumber\\
&&+\frac{ck_B}{e}\left(n_z\frac{\partial T_e}{\partial R}-n_R\frac{\partial T_e}{\partial z}\right)
\left[\ln P_{e,d} - \ln P_{e,u}\right]\equiv\nonumber\\
&&a_d-a_u+b_d-b_u = 0,
\label{eq:RH_Verify_1}
\eea
where we've defined ``Advection'' terms $a_{u,d}$ and ``Biermann'' terms $b_{u,d}$
by
\bea
a_d&\equiv&(D-u_d)B_d\label{eq:A_d}\\
a_u&\equiv&(D-u_u)B_u\label{eq:A_u}\\
b_d&\equiv&\frac{ck_B}{e}\left(n_z\frac{\partial T_e}{\partial R}-n_R\frac{\partial T_e}{\partial z}\right)
\ln P_d\label{eq:B_d}\\
b_u&\equiv&\frac{ck_B}{e}\left(n_z\frac{\partial T_e}{\partial R}-n_R\frac{\partial T_e}{\partial z}\right)
\ln P_u.\label{eq:B_u}
\eea

The sum of terms in Eq.~(\ref{eq:RH_Verify_1}) is required to be zero.  In a discretized
numerical PDE integration, this really means that the terms must cancel up to
some truncation or rounding precision, which is expressed relative to the largest
of the magnitudes in play.  We therefore define the ``shock condition parameter'' $C$
by
\beq
C\equiv\frac{a_d-a_u+b_d-b_u}{\mbox{max}(|a_d|,|a_u|,|b_d|,|b_u|)}.\label{eq:Shock_Condition_Parameter}
\eeq

We calculate the value of $C$ at cells along the shock front, at each of our 
four resolution levels, for both the correct flux and the ``naive'' flux 
implementations of the Biermann effect.  We display cumulative distributions 
of $C$ in the top panels of Figure~\ref{Figure:NonSpherical}.  It is evident 
from these figures that the distribution of $C$ is in fact centered near 
zero for both flux implementations.  The width of the distributions behave 
very differently, however.  In the case of the correct Biermann flux 
implementation, the distributions get narrower with each refinement of 
resolution, providing some evidence of convergence with resolution to the 
expected result.  In the case of the naive flux implementation, there is no 
such evidence of convergence. The same observations can be made of the plots 
in the middle panels of Figure~\ref{Figure:NonSpherical}, which summarize 
the means and sample standard deviations of the $C$-distributions, as a 
function of resolution.  The convergence of the correct flux implementation, 
and the convergence failure of the naive implementation, are clear in these 
figures.

The bottom panels of Figure~\ref{Figure:NonSpherical} show the evolution of 
total magnetic energy in the domain as a function of time, for the different 
resolutions and for the two flux implementations.  The correct flux 
implementation appears to be converging (although not in any strong sense 
converged) by this measure, even at late time, whereas any evidence of 
convergence in magnetic field energy is simply lacking for the naive flux 
implementation.

\subsubsection{The Resistive Magnetic Precursor}

In this set of simulations, we maintain the simulation parameters 
described in \S\ref{subsubsec:Ellipsoidal_verification}, but also turn on 
the resistivity $\eta$ to a finite positive value, $\eta=7.8\times 10^{5}$~s
$^{-1}$, chosen in conjunction with the other parameters to yield an 
easily-discernible exponentially-decaying resistive precursor described by 
Eq.~(\ref {eq:precursor_upstream_solution}).  We repeat the simulation 
strategy of \S\ref{subsubsec:Ellipsoidal_verification}, letting the 
simulation advance to $t=4.86\times 10^{-8}$~s with no field generation, 
then re-starting it and advancing to a simulation time $t=3\times10^{-7}$~s 
with field generation (using the correct flux formulation) turned on. This 
advance time is adequate for the establishment of the expected precursor 
structure, as we demonstrate below.

The left panel of Figure~\ref{Figure:Resistive} shows a colormap of the 
distribution of magnetic field strength across the domain. The shock 
location is shown by the solid red line, and the superposed vectors indicate the 
shock-normal direction.  The magnetic field is evidently smoothed out by the 
resistivity, as can be seen by a direct comparison with the bottom-left 
panel of Figure~\ref{Figure:NonSpherical_B_Pseudocolor}.  The precursor is
already evident in this figure, as the substantial amount of magnetic field
that has ``leaked'' ahead of the shock.

\begin{figure}[h]
\begin{center}
\includegraphics[width=3.25in]{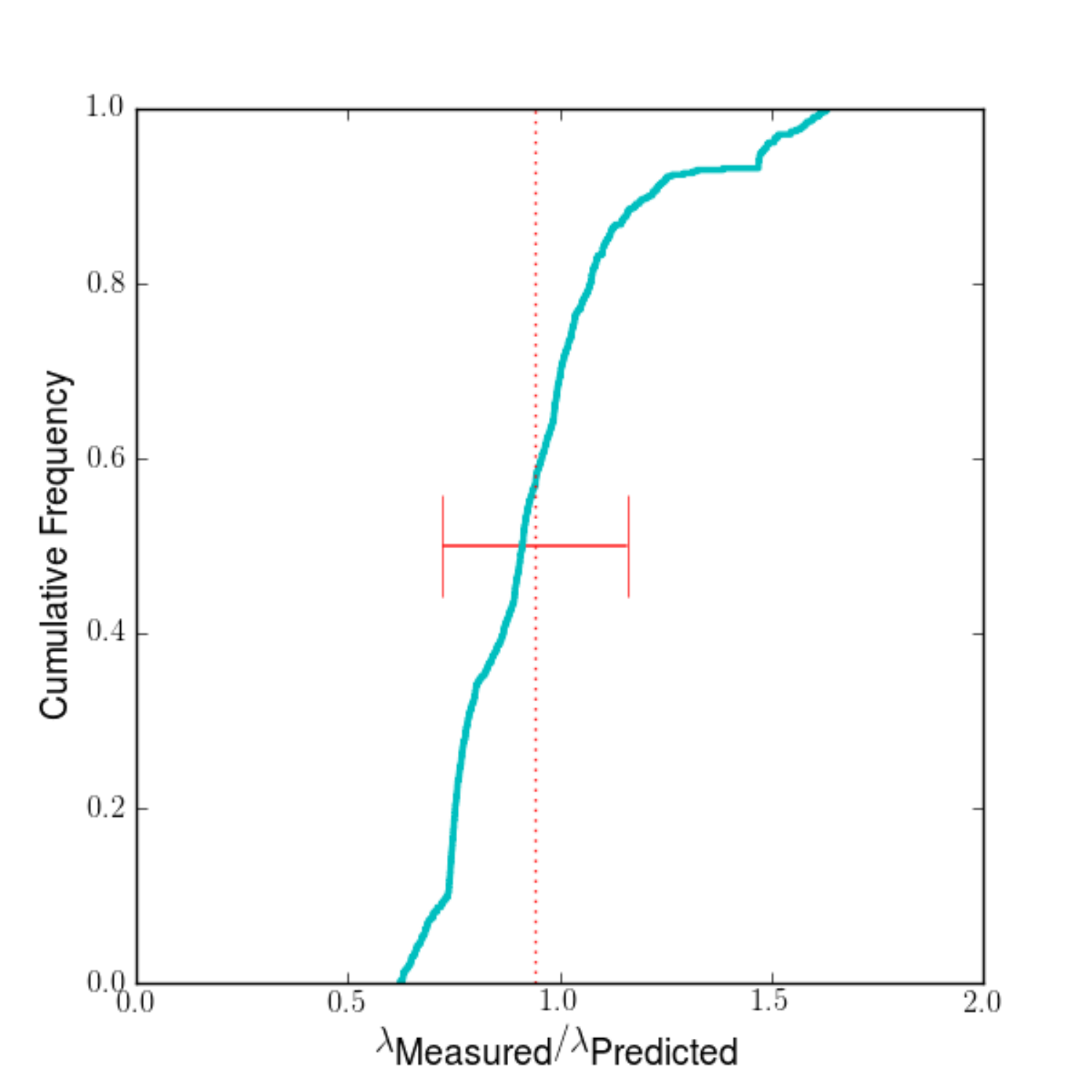}
\end{center}
\caption{Cumulative distribution of the measured magnetic precursor decay length,
normalized to the predicted length,
at 849 points along the shock surface.  The mean is shown by the dotted line,
while the standard deviation is illustrated by the errorbar.
\label{Figure:Resistive_Verification}}
\end{figure}

The diagonal white dashed shock-crossing in the left panel of Figure ~\ref{Figure:Resistive} illustrates the line along which data 
was extracted to produce the right panel of Figure~\ref{Figure:Resistive}.
This figure shows the magnetic field values plotted along that line.  The
vertical solid red line at $X=0$ marks the location of the shock.  The exponential
decay of the field is easily visible in this figure.

At each shock location, the value of $\eta$ can be combined with the local 
shock velocity to calculate $\lambda_{\mbox{Predicted}}$, the expected decay 
length of the precursor, according to Eq.~(\ref{eq:lambda_B}). At the same 
time, the actual decay length $\lambda_{\mbox{Measured}}$ can be measured 
directly by comparing the field strength at two suitably-selected distances 
along the local shock normal.  We have performed both calculations at each 
shock location, and compared them.  The results are displayed in Figure~\ref 
{Figure:Resistive_Verification}, where we plot the cumulative distribution 
of $\lambda_{\mbox{Measured}}/\lambda_{\mbox{Predicted}}$.  The distribution
appears to be consistent with a mean value of 1, as expected, with some
scatter.  The scatter is not surprising, since the expression in 
Eq.~(\ref{eq:lambda_B}) for $\lambda_{\mbox{Predicted}}$ is an approximation,
assuming, as it does, propagation of magnetic field from a planar shock.
Since the shock is, of course, not planar, the value of the Biermann generation
rate changes appreciably across the shock over distances not hugely different
from $\lambda_{\mbox{Predicted}}$ itself.  Under the circumstances, then, the
level of agreement between observation and prediction is satisfactory.

\begin{figure*}[t]
\begin{center}
\begin{minipage}[b]{3.4in}
\includegraphics[width=3.55in,viewport=0 70 1024 800,clip=]{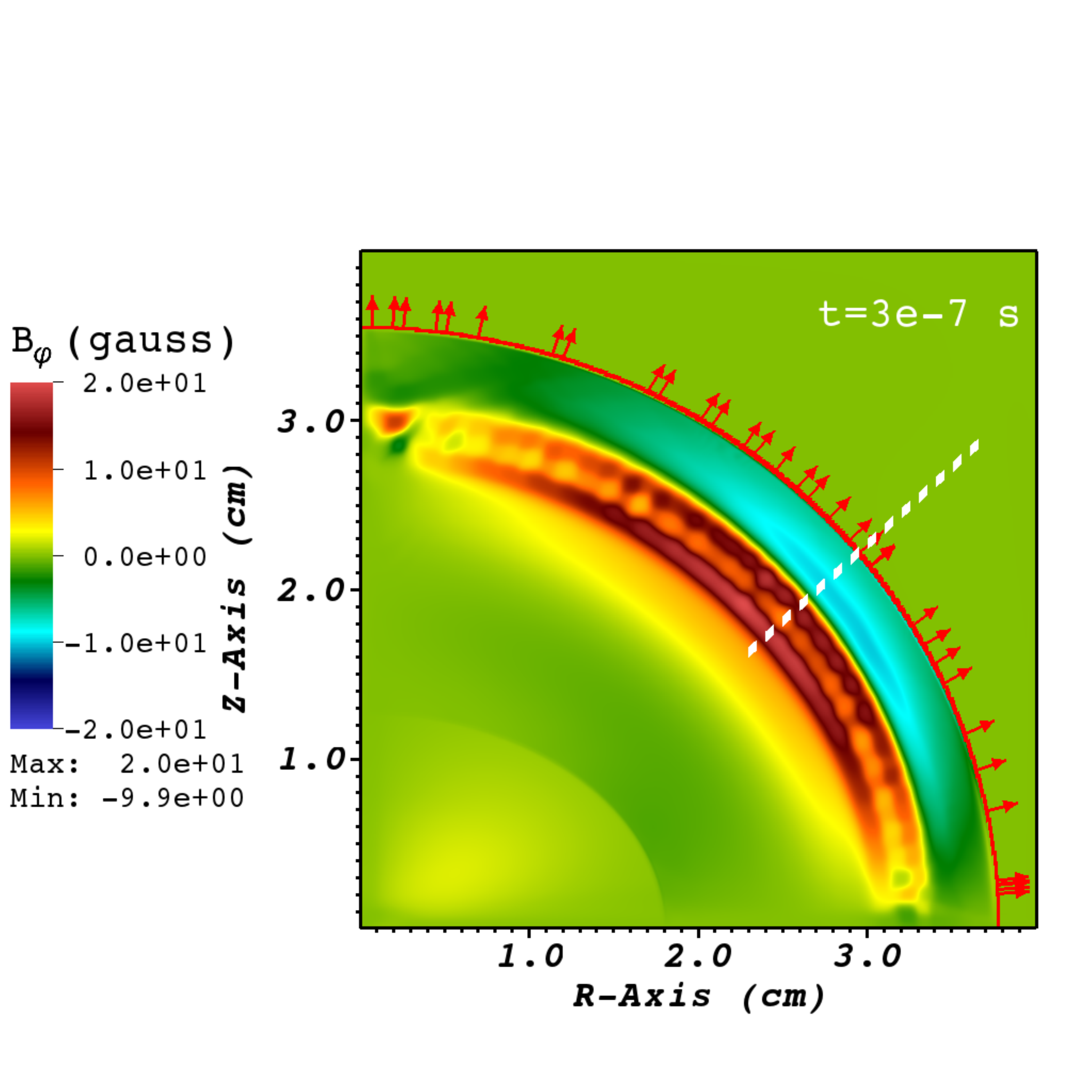}
\end{minipage}
\begin{minipage}[b]{3.4in}
\includegraphics[width=3.0in]{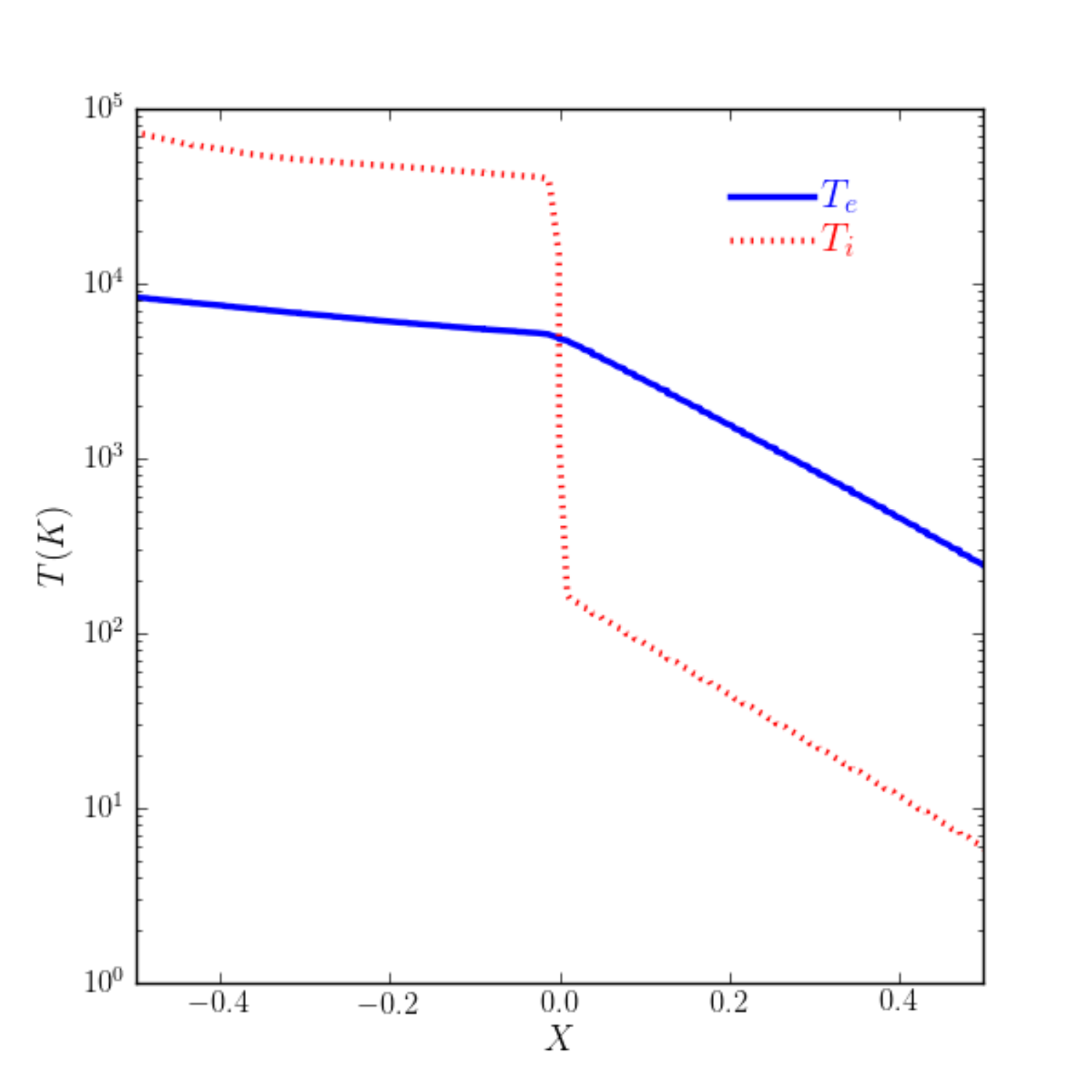}
\end{minipage}
\includegraphics[width=3.0in]{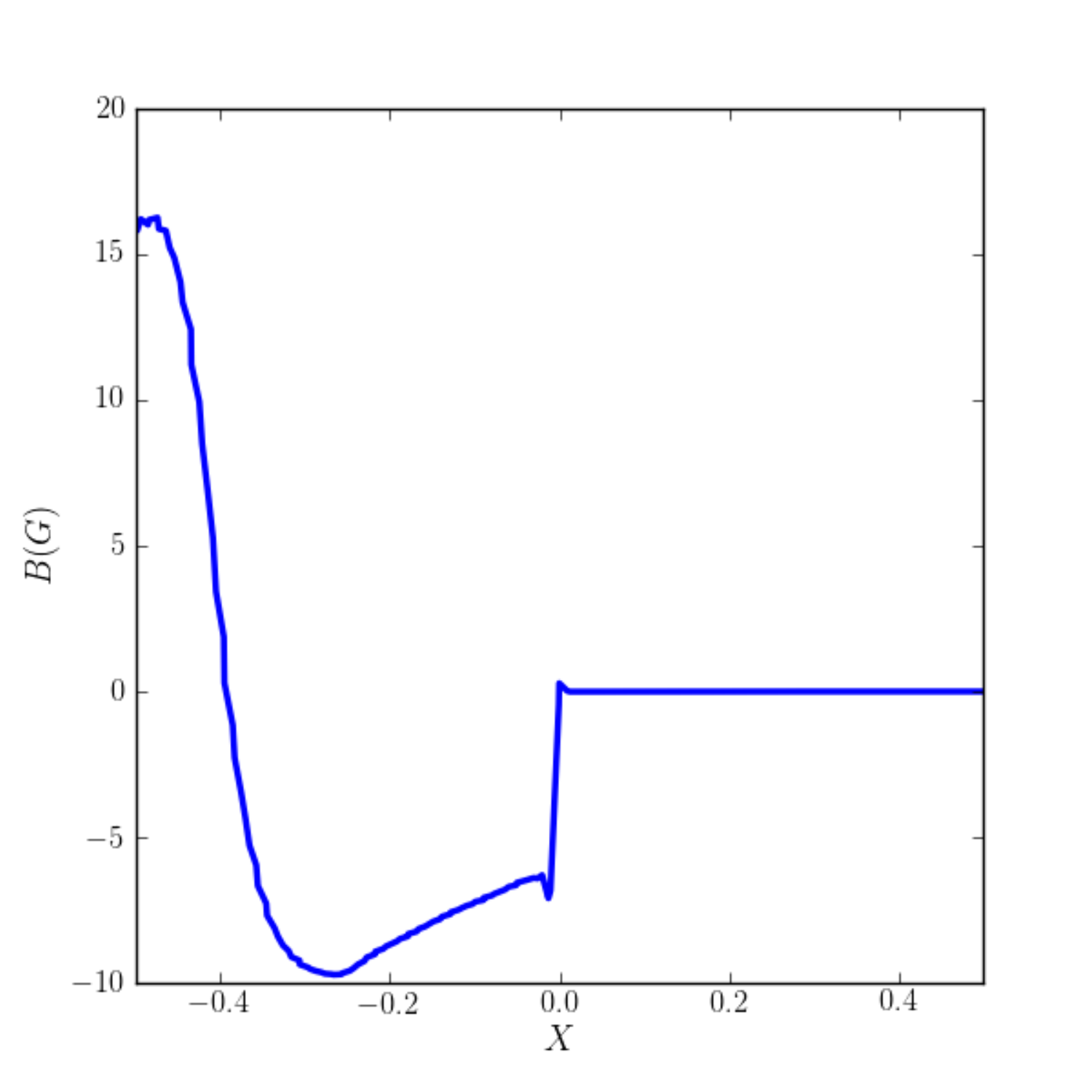}
\includegraphics[width=3.0in]{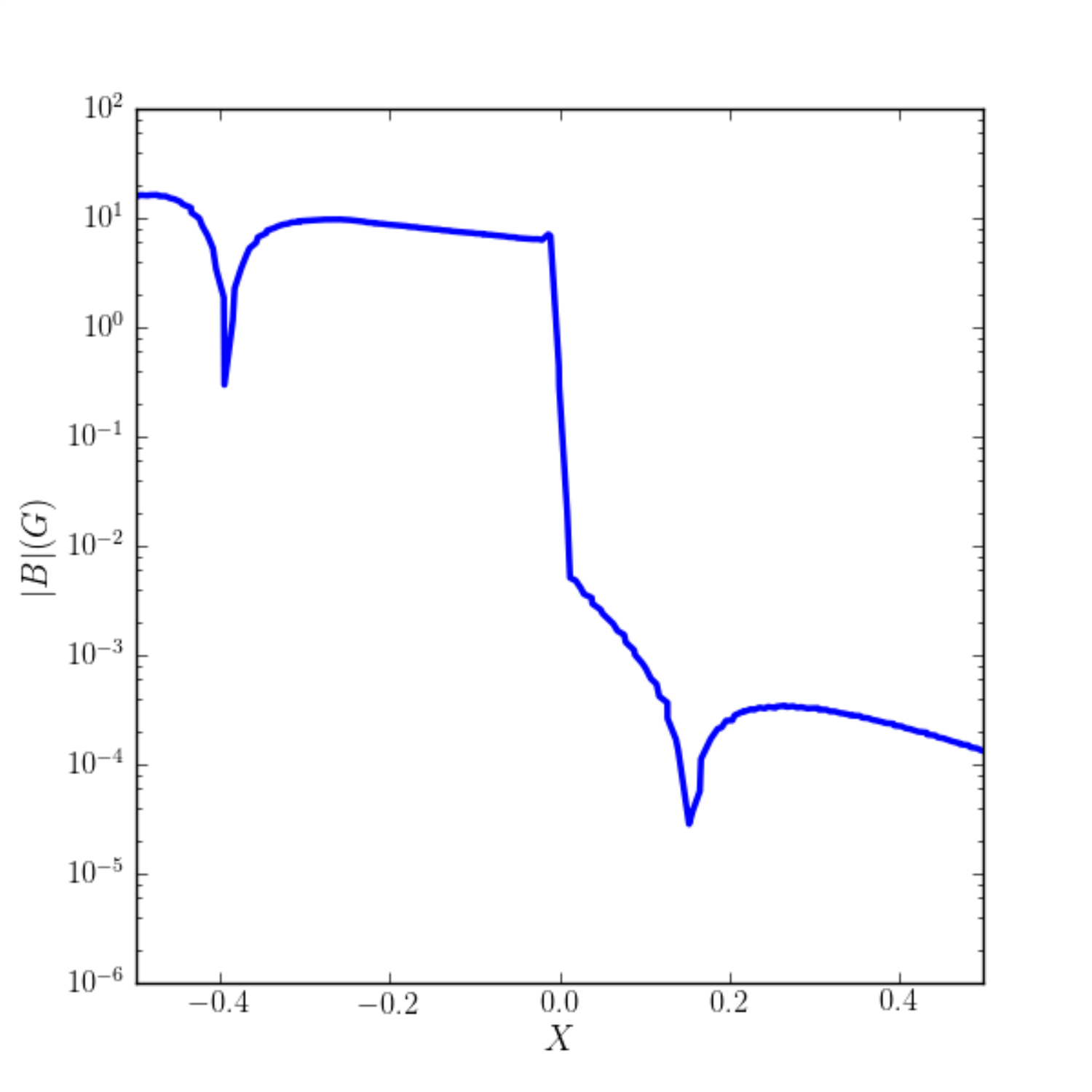}
\end{center}
\caption{Ellipsoidal shock, with 10$\times$ greater thermal conductivity. 
Top Left: Final magnetic field distribution. The solid red line shows the shock
location, while the arrows show the direction of the shock normal. The white dashed shock-crossing line 
illustrates the location of the data displayed in the remaining panels; Top 
Right: Electron and ion temperatures along the white dashed shock-crossing line in the 
previous panel.  The shock location and thermal precursor are evident; 
Bottom Left: $B$ along the shock-crossing line; Bottom Right: Semilog plot of $|B|$
along shock-crossing line.
\label{Figure:Thermal_Precursor}}
\end{figure*}

\subsubsection{The Thermal Magnetic Precursor}\label
{subsubsec:thermal_magnetic_sim} In our final simulation, we explore the 
properties of the thermal magnetic precursor by returning to a non-resistive 
simulation, similar to the ones discussed in \S\ref
{subsubsec:Ellipsoidal_verification}, and differing from them only in that 
$\kappa_e$ is ten times larger:  
$\kappa_e=1.0\times 10^{14}$~erg~K$^{-1}$~cm $^{-1}$~s$^{-1}$.  This broadens
the thermal precursor zone to about 0.2~cm, making it easier to discern.

The results at the final time step are illustrated in Figure~\ref
{Figure:Thermal_Precursor}. The top-left panel shows the distribution of
magnetic field in the domain.  The solid red line in the figure shows the location
of the shock, while the arrows show the direction of the shock normal.  In the top-right panel we see the electron
and ion temperatures along the shock-crossing line shown in the previous
panel.  The shock location coincides with the sharp drop in $T_i$, while the
thermal conduction precursor zone, which has an exponential structure for
the constant conductivity model used here, may be clearly seen in the
upstream behavior of $T_e$.

The magnetic field structure along the shock-crossing line is shown in the 
bottom-left panel of the figure. The precursor is difficult to see in this 
plot, so the bottom-right panel shows $|B|$ on a semi-log plot.  The 
magnetic precursor structure is seen in this figure.  It clearly has a much 
lower field intensity than the fields generated by the Biermann effect at 
the shock.  We emphasize again, however, that in part this is due to the 
highly simplified constant-conductivity model adopted for this work.  A true 
Spitzer-type conductivity, with a $T_e^{5/2}$ dependence, would create much 
sharper gradients at the upstream termination of the precursor zone, potentially
generating more intense fields -- relative to those generated at the shock --
than those shown here.

\section{Summary and Discussion}\label{sec:discussion}

To summarize our findings:  Using the kinetic theory of plasma shocks, we 
have given a description of the Biermann effect within ion viscous shocks.
Using this description, we have shown that the convergence 
failures in the presence of shocks of MHD codes implementing the Biermann 
effect is not traceable either to a mathematical mis-specification of the 
Biermann term -- which is well-defined despite appearances -- or to a 
failure of the Biermann MHD source term to correctly approximate the 
expected physics.  The failure is instead due to naive discretization, a
condition that we have shown is curable by exploiting the continuity of
the electron temperature $T_e$.

We have described a convergent algorithm that incorporates the Biermann 
effect within shocks in numerical MHD, and implemented it in the FLASH code.  
We have developed and used verification tests to 
demonstrate formal convergence using a null solution from spherical shock, 
and to verify predictions for physical conditions near a shock using an 
ellipsoidal shock. Comparisons of the new algorithm, which provides a correct
and accurate treatment of the Biermann effect within shocks, with the previous, 
naive one exhibit clearly desirable physical and numerical properties present 
in the new algorithm that were previously wanting.

We have noticed, described, and simulated two previously unrecognized 
physical effects: a resistive magnetic precursor that leads the shock in the 
presence of non-zero resistivity, and that is analogous to the well-known 
thermal precursor caused by the presence of finite electron thermal 
conduction; and a thermal magnetic precursor, produced by plasma motions in 
the electron thermal conduction precursor.  We have estimated the expected 
magnitude of both effects in conditions encountered in laboratory 
experiments at laser facilities, and shown that the characteristic length of 
the two effects can be made macroscopic.  In particular, the resistive 
magnetic precursor is physically measurable in an experimental setup that 
measures both the position of the shock front and a time-series of the 
intensity of the magnetic field at some location initially ahead of the 
shock.  Such an experiment would have to somehow ensure that the upstream 
fluid remains unheated by shock-generated radiation and by laser absorption, 
so as to keep the value of the resistivity high enough to sustain a 
macroscopically-scaled precursor length. The thermal magnetic precursor may 
be more challenging to observe, as it is intrinsically weaker than the 
resistive precursor (relative to the field intensity generated at the 
shock).  Further studies with Spitzer-type electron thermal conductivity are 
needed to determine whether sharp gradients at the upstream termination of 
the electron thermal precursor can give rise to sufficiently intense 
magnetic field strengths to be experimentally measurable.

We emphasize again that the requirement for convergence in $\bB$ of the new 
Biermann effect algorithm is that the electron thermal precursor of the 
shock should be resolved, while the requirement for convergence in magnetic 
energy of the new algorithm is that the resistive precursor should be 
resolved.  Of these two requirements, the first is more essential, because 
neglect of the contribution of the Biermann effect to the flux of magnetic 
energy is frequently a valid approximation.  Demanding resolution of the 
electron thermal precursor is not a trivial requirement in realistic 
simulations.  It certainly mandates some kind of Adaptive Mesh Refinement 
(AMR) strategy to resolve the region near the shock.  But irrespective of 
AMR-related efficiencies, merely resolving that length scale can constrain 
the code to taking very short time steps due to the Courant-Friedrichs-Lewy 
(CFL) stability condition, unless an implicit advancement scheme is 
implemented for MHD.  This issue is the subject of current study, but it 
lies beyond the scope of the present paper.

With the discussion of the modification due to the Biermann effect of the 
Rankine-Hugoniot jump condition on $B$ -- Eq.~(\ref{eq:Biermann_RH}) and 
text thereabout -- it now seems worth returning to the question of vorticity 
$\bm{\omega}$, and the beguiling Equation~(\ref{eq:Mag_vs_Vorticity}) that 
suggests its proportionality to $\bB$. Recall that much hangs on the jump 
conditions -- if the jump in $\bm{\omega}$ maintains the same 
proportionality to the jump in $\bB$ as do their respective evolution 
equations, then, subject to some cavils about boundary conditions, $\bB$ is 
simply $\bm{\omega}$, at least until resistivity and/or viscosity manifest 
themselves, and in effect, unmagnetized gasdynamics contains within it the magnetic
degrees of freedom of MHD, encoded in derivatives of the velocity field.

It can be seen more clearly now why this implausible assertion is in fact 
false. In the first place, $\bB$  and $\bm{\omega}$ are rather different 
types of quantities from the point of view of kinetic theory, in that $\bB$ 
is perfectly well-defined at kinetic scales -- such as inside an ion viscous 
shock sheath -- whereas $\bm{\omega}$ cannot even be defined consistently in 
regions where the fluid picture breaks down.  The jump condition on $\bB$ is 
a straightforward consequence of the induction equation, and is essentially 
kinematic in origin, as are the other Rankine-Hugoniot conditions.  The jump 
condition on vorticity, on the other hand, cannot be inferred from the 
dynamical PDE equations alone, but depends in an essential manner on the 
equation of state \citep{kevlahan1997}. It would be possible, in principle, 
to infer a ``flux'' of $\bm{\omega}$ from Eq.~(\ref{eq:vorticity}), and to 
construct a ``jump condition'' using that flux in the Rankine-Hugoniot 
conditions, Eq.~(\ref{eq:RH}).  That jump condition would disagree with the 
one obtained in \citep{kevlahan1997}, as it is manifestly independent of the 
EOS.  The EOS dependence of the jump condition on vorticity implies that 
even if there exists an EOS according to which the jump condition on 
$\bm{\omega}$ preserves the required proportionality to the Biermann-laden 
jump in $\bB$, that proportionality would certainly not be preserved for any 
other EOS.  

In general, then, the passage of a shock certainly spoils Eq.~(\ref 
{eq:Mag_vs_Vorticity}).  This should be no surprise, as the manipulations of 
the hydrodynamic equations of motion required to arrive at Eq.~(\ref 
{eq:vorticity}) constitute precisely the sorts of prestidigitation with 
vector derivatives that break down at fluid discontinuities \citep[see p. 
216 of] [for example]{leveque2002}. While Eq.~(\ref{eq:vorticity}) may be 
derived from the momentum equation on either side of the shock, the 
connection between $\bB$ and $\bm{\omega}$ is certain to be lost at the 
shock itself, well before resistive and viscous effects can assert 
themselves.  We conclude that the vorticity connection is not a useful
tool for studying magnetogeneration by the Biermann effect in the presence
of shocks.

In concluding, it is worth pointing out again that since shocks are not the 
only weak-solution discontinuities that arise in multi-material MHD 
simulations, they are not the only locations where potential convergence 
failures are corrected by the new algorithm.  In particular, contact 
discontinuities, material discontinuities, and ionization fronts all present 
potential problems for the naive Biermann algorithm, since they all 
represent locations where $P_e$ changes discontinuously (despite the 
continuity of total pressure $P$ at such locations), and are therefore all 
sites where the Biermann effect can be expected to generate magnetic field. 
They all represent physical situations in which the new algorithm provides a 
correct treatment of the Biermann effect.

\acknowledgements 
We would like to thank Gianluca Gregori for discussions that helped to focus 
this work on the relevant physical issues.  We would also like to thank the 
anonymous referee for encouraging us to discuss more fully the astrophysical
implications of this work; and Andrey Kravtsov, Brian O'Shea, and John ZuHone 
for correspondence and conversations about the application of this work to
the creation of magnetic fields by the Biermann effect in galaxy clusters.
This work was supported in part by the National Science Foundation under 
grant AST-0909132; and in part by the U.S. DOE NNSA ASC through the Argonne 
Institute for Computing in Science under field work proposal 57789. The 
software used in this work was developed in part by funding from the U.S. DOE 
NNSA-ASC and OS-OASCR to the Flash Center for Computational Science at the 
University of Chicago.

\appendix

\section{Shock Detection}\label{appendix:Shock_Detection}

We describe here our shock detection algorithm, which we used in the 
verification of the Biermann effect at shocks.  This algorithm has some 
benefits over shock-detection algorithms such as the ones described in
\citet{balsara1999,Miniati2000,Ryu2003}, in that it furnishes an estimate of the shock speed directly 
from a single time slice of spatial data, weights the mass, momentum, and 
energy components of the Rankine-Hugoniot relations equally, yields a 
single-cell-wide shock interface, verifies that local characteristics
converge on the shock surface, and does not impose arbitrary thresholds on
physical quantities such as compression ratios or Mach numbers.

We start with a mesh of cells containing fluid values (for present purposes it
is immaterial whether these are cell averages or point values).  The algorithm
seeks a set of ``shocked'' cells satisfying the following criteria:

\begin{itemize} 

\item The Shocked Surface Is One Cell Wide: Each shocked cell is the 
location of a maximum of $|\nabla P|$ in the direction of $\nabla P$, where 
$P$ is the total fluid pressure;\label{item:MaxGradP}

\item Fluid Quantities Change Correctly: The neighborhood of each shocked 
cell exhibits compression, acceleration, and heating in the direction of 
$\nabla P$;\label{item:CompAccHeat}

\item Rankine-Hugoniot Conditions Obtain: The full Rankine-Hugoniot 
conditions on mass, momentum, and energy hold between the fluid upstream and 
downstream of each shocked cell, where ``upstream'' means in the direction 
$\bn\equiv-\nabla P/|\nabla P|$;\label{item:RH}

\item Characteristics Converge On The Shock Surface: The characteristic 
convergence criterion holds at each shocked cell:  the shock is supersonic 
upstream, and subsonic downstream.\label{item:CharConv} 

\end{itemize}

The result is an efficient and reliable algorithm that among other things 
yields an accurate estimate of the shock speed, which is essential to our 
verification work on the Biermann effect.  Note that we neglect the 
induction equation in the shock detection algorithm, and operate using only 
the material pressure -- not the total pressure, which includes the magnetic 
pressure.  For weakly magnetized plasmas, such as the ones we consider in 
this paper, this creates no difficulty in identifying the correct shock 
surface.  Some generalization would be required for significantly magnetized 
plasmas, particularly to the third and fourth conditions above.

We now describe in somewhat greater detail the algorithm outlined above.

\subsection{The Shocked Surface Is One Cell Wide}

The first condition above is tantamount to insisting that the
shocked cell be sited at a position where the gradient of $P$ is steepest.  In
general, volume-based hydrodynamic advance schemes spread shocks over several
cells.  If we were not to impose this condition, we would obtain a thickened
shell of shocked cells, which would
complicate the criterion (specified below) for classifying neighboring cells
as adjoining the shock upstream or downstream.

The condition is very easy to enforce:  in the immediate neighborhood of the 
cell being tested for ``shockedness'', we check the adjoining cells closest to 
the direction of the normal vector $\bn$ and its mirror image $-\bn$.  This 
is done by forming the vector of integers $\Delta\bm{i}\equiv\mbox{NINT}[\bn/\mbox{max}(n_1,n_2,n_3)]$
(where the NINT function ascribes the nearest integer to each component of
its real vector argument) and 
adding it to the cell discrete index vector $\bm{i}$, to reach the cells at 
$\bm{i}\pm\Delta\bm{i}$. If a stencil-based estimate of $|\nabla P|$ is 
greater in the candidate cell than in the two so-chosen adjoining cells, 
then the condition is passed successfully.

\subsection{Fluid Quantities Change Correctly}

An easy sanity check for ``shockedness'' of a cell is that cells downstream 
should unfailingly be (a) compressed, (b) accelerated (in the $+\bn$ -direction), 
and (c) heated, with respect to cells upstream.  

We introduce a shock-width parameter $h$, such that the shock-adjoining 
cells ``upstream'' and ``downstream'' of the shocked cell are respectively located at $\pm 
h\times\Delta\times\bn$ relative to the cell under study, where $\Delta$ is 
the length of a cell side.  For FLASH with the HLL Riemann solver \citep 
{toro2009} and 2nd-order reconstruction, an appropriate value of $h$ is 
1.5.  We then simply verify that the fluid variables $\rho$, $\bu$, and $e_T$
(the total thermal energy) satisfy the conditions $\left[\rho\right]^d_u>0$, 
$\left[\bu\cdot\bn\right]^d_u>0$, and $\left[e_{T}\right]^d_u>0$.

\subsection{Rankine-Hugoniot Conditions Obtain}

The Rankine-Hugoniot (R-H) conditions, which express flux conservation in the frame
of the shock, have the form
\beq
D\times(\bm{\Phi}_d-\bm{\Phi}_u)=
\bm{F}(\bm{\Phi}_d)-\bm{F}(\bm{\Phi}_u).
\label{eq:RH2}
\eeq
Here, $\bm\Phi$ is the state vector of conserved field quantities, 
$\bm\Phi^T\equiv[\rho,\rho\bu\cdot\bn,\rho{\cal E}]$, that is, the mass, normal
momentum, and energy densities. $\bm{F}$ denotes the vector
of fluxes corresponding to $\bm{\Phi}$.  The subscripts ``d'' and ``u'' denote
``downstream'' and ``upstream'' states, respectively \citep{toro2009}.

We fit R-H conditions to the upstream/downstream data defined using the shock-width
parameter $h$ above.  In this fit, the shock speed $D$ is a free parameter to be
adjusted to minimize a fit residual.  Given a positive-definite inner product $\left[\Phi_1,\Phi_2\right]$ on
the vector state space in Eq.~(\ref{eq:RH2}), we may define the normalized residual $R^2$ as
\beq 
R^2\equiv \frac{\left[\left(D\Delta\bm{\Phi}-\Delta\bm{F}\right),
\left(D\Delta\bm{\Phi}-\Delta\bm{F}\right)\right]}{\left[\Delta\bm{F},\Delta\bm{F}\right]},
\eeq
where $\Delta\bm{\Phi}\equiv\bm{\Phi}_d-\bm{\Phi}_u$ and 
$\Delta\bm{F}\equiv\bm{F}(\bm{\Phi}_d)-\bm{F}(\bm{\Phi}_u)$.  We may then easily
minimize $R^2$ with respect to $D$, obtaining
\bea
D_{Min}&=&\frac{\left[\Delta\bm{\Phi},\Delta\bm{F}\right]}
{\left[\Delta\bm{\Phi},\Delta\bm{\Phi}\right]}\label{eq:Best_Fit_D}\\
R^2_{Min}&=&1-\frac{\left[\Delta\bm{F},\Delta\bm{\Phi}\right]^2}
{\left[\Delta\bm{F},\Delta\bm{F}\right]\left[\Delta\bm{\Phi},\Delta\bm{\Phi}\right]}.
\label{eq:Min_Residual}
\eea

Given these definitions, we consider that the R-H conditions are satisfied if
$R^2_{Min}$ is less than some small threshold value.

It remains to define the inner product $[\cdot,\cdot]$ used in these expressions.  It is
obvious that the naive unweighted sum-of-squares of the components of vectors
such as $\Delta\bm{\Phi}$ is not an acceptable inner product, as it is dimensionally
senseless.  We require at a minimum a positive-definite metric that brings all
vector components to the same physical dimensions, so we don't wind up adding
mass densities to energy densities, and so on.  An additional desirable requirement
is that all three components of Eq.~(\ref{eq:RH2}) should contribute similar magnitudes
to Eqs.~(\ref{eq:Best_Fit_D}) and (\ref{eq:Min_Residual}), so that they are all
weighted equally in the fit.  

A simple metric that accomplishes these requirements
may be build using the local sound speed $c_s$ of the candidate shocked cell:
\beq
\left[\bm{\Phi}_1,\bm{\Phi}_2\right]\equiv{\bm{\Phi}_1}^T\bm{M}\bm{\Phi}_2,
\label{eq:Inner_Product}
\eeq
where
\beq
\bm{M}\equiv\left[
\begin{array}{lll}
1&0&0\\
0&{c_s}^{-2}&0\\
0&0&{c_s}^{-4}\\
\end{array}
\right].
\label{eq:Metric}
\eeq
This choice produces compatible dimensions and comparable magnitudes, 
because in and downstream of the shock we have 
$c_s\sim (k_B  T/m_i)^{1/2}\sim D$, so that in the frame of the shock, the mass 
flux is $\rho D\sim\rho c_s$, the momentum flux has magnitude $P\sim\rho {c_s}^2$,
and the energy flux has magnitude $\rho D^3\sim\rho {c_s}^3$.

We adopt this metric in computing Eqs.~(\ref{eq:Best_Fit_D}) and (\ref{eq:Min_Residual}).
We find that a threshold of $R^2_{Min}<0.01$ is adequate for identifying cells
satisfying the R-H conditions with few false-positives and false-negatives, in
the simulations exhibited in this paper.

\subsection{Characteristics Converge On The Shock Surface}

Our final criterion is easily stated and checked: shocks are supersonic 
upstream, and subsonic downstream, so that characteristics from the family 
implicated in the shock converge on the shock \citep[see][pp. 141-146]{courant1948}.  
This is an essential stability criterion for the solution.  Using the shock speed
$D_{Min}$ calculated while fitting the R-H conditions to the data, this condition
is simply
\beq
\bu_d\cdot\bn+c_{s,d}>D_{Min}>\bu_u\cdot\bn+c_{s,u},\label{eq:ConvChar}
\eeq
where $c_{s,d}$ and $c_{s,u}$ are the downstream and upstream sound speeds,
respectively, as determined with respect to the shock width $h$ defined above.

\bibliography{biermann}

\end{document}